\documentclass[final]{agujournal}

\usepackage[english]{babel}
\usepackage[T1]{fontenc}
\usepackage[utf8]{inputenc}
\usepackage{array}
\usepackage{multirow}
\usepackage{amsmath}
\usepackage{graphicx}
\usepackage{rotating}

\usepackage{subcaption}
\usepackage{url}
\usepackage{bbm}
\usepackage{dsfont}

\usepackage{babel}

\journalname{<>}

\begin{document}
%
%


\title{Effects of Spatiotemporal Upscaling on Predictions of Reactive Transport in Porous Media}

%
%




\authors{Farzaneh Rajabi\affil{1} }


\affiliation{1}{Department of Energy Resources Engineering, Stanford University, Stanford, CA, USA}




\correspondingauthor{Farzaneh Rajabi}{frajabi@stanford.edu}




\begin{keypoints}
\item Introduction of the concept of spatiotemporal upscaling in the context of homogenization by multiple-scale expansions. 
\item Impact of time-dependent forcings and boundary conditions on macroscopic reactive transport in porous media.
\item The dynamics at the continuum scale is strongly influenced by the interplay between signal frequency at the boundary and transport processes at the pore level.
\end{keypoints}

%
%


\begin{abstract}
The typical temporal resolution used in modern simulations significantly exceeds characteristic time scales at which the system is driven. This is especially so when systems are simulated over\deleted{ultra-long} time-scales \added{that are much longer than the typical temporal scales of forcing factors}. We investigate the impact of space-time upscaling on reactive transport in porous media driven by time-dependent boundary conditions  whose \deleted{frequency} \added{characteristic time scale} is much \deleted{larger} \added{smaller} than \deleted{the characteristic time scale} \added{that} at which transport is studied or observed at the macroscopic level. The focus is on transport of a reactive solute undergoing diffusion, advection and heterogeneous reaction on the solid grains boundaries.   We first  introduce a concept of spatiotemporal upscaling in the context of homogenization by multiple-scale expansions, and demonstrate the impact of time-dependent forcings and boundary conditions on macroscopic reactive transport. We then derive the macroscopic equation as well as the corresponding applicability conditions based on the order of magnitude of the P\'{e}clet and Damk\"{o}hler  dimensionless numbers. Finally, we   demonstrate that the dynamics at the continuum scale is strongly influenced by the interplay between signal frequency at the boundary and transport processes at the pore level.

%

\end{abstract}

%
%

%


%
%
%
%

\section{Introduction}

\deleted{Despite significant progress, accurate quantitative predictions of subsurface transport of highly reactive fluids remains a formidable challenge. Current numerical models suffer from significant predictive uncertainty, which undermines our ability to estimate future impact of, and the risks associated with, anthropogenic stressors on the environment. That is because subsurface flow and transport take place in complex highly hierarchical heterogeneous environments, and exhibit nonlinear dynamics and often lack spatiotemporal scale separation.}
The choice of an appropriate level of hydrogeologic model complexity continues to be a challenge. \added{That is because subsurface flow and transport take place in complex highly hierarchical heterogeneous environments, and exhibit nonlinear dynamics and often lack spatiotemporal scale separation \citep{tartakovsky-2013-assessment}.} The constant tension between fundamental understanding and predictive science on the one hand, and the need to provide science-informed engineering-based solutions to practitioners, on the other,  is part of an ongoing debate on the role of hydrologic models \citep[e.g.][]{miller-2013-Numerical}. A physics-based model development follows a bottom-up approach which, through rigorous upscaling techniques, allows one to construct effective medium representations of fine-scale processes with different degrees of coupling and complexity \citep[e.g.][]{wood2013volume,helming-2013-model}. Yet, current model deployment is generally based on established engineering practices and often relies on `simpler' classical \deleted{single-point closure}  \added{local} continuum descriptions with limited predictive capabilities.

 The development of  multiscale, multiphysics models aims at  filling this scale gap and at addressing the limited applicability of classical local macroscopic models \citep{Auriault-1991-heteronenous,auriault1995taylor,mikelic2006rigorous}. Originated in the physics literature, multiscale methods were developed to couple particle to continuum solvers \citep{Wadsworth-1990-One,Hadjiconstantinou-1997-heterogeneous,Abraham-1998-Spanning,Tiwari-1998-coupling,Shenoy-1999-adaptive,Flekkoy-2000-hybrid,alexander-2002-algorithm,Alexander-2005-noise}. Multiphysics domain-decomposition approaches \citep{Malgo-2002-mortar,arbogast-2007-multiscale,ganis2014global}, combined with multiscale concepts, led to the development  of  multiphysics, multiscale capabilities  to address the multiscale nature of transport in the subsurface \citep{tartakovsky2008hybrid,mehmani2012multiblock,roubinet2013hybrid,Bogers-2013-multiscale,mehmani2014bridging, yousefzadeh2020numerical,Taverniers-2017-tightly}. The proposed methods predominantly focus on tackling partial or total lack of scale separation due to spatial heterogeneity, and are often based on spatial upscaling to construct coupling conditions between representations at different scales.

Upscaling methods enable one to formally establish a link between fine-scale (e.g. pore-scale)  and   observation-scale/macroscopic processes. Spatial upscaling methods include volume averaging \deleted{averaging} \citep[e.g.,][]{wood2003volume,wood2009role,whitaker2013method,wood2013volume}  \deleted{and its modifications} \added{and thermodynamically constrained averaging theory} \added{\citep{Gray-2005-Thermodynamically,gray-2014-intro}}, the method of moments \citep{taylor1953dispersion,brenner1980dispersion,shapiro1988dispersion}, homogenization via multiple-scale expansions \citep[][e.g.,]{bensoussan1978asymptotic,hornung1994reactive,allaire2010homogenization,hornung2012homogenization},  and pore network models \citep{acharya2005transport}. \cite{cushman2002primer} provides a review of different upscaling methods. Comparative studies discuss differences and similarities of various upscaling techniques \citep[e.g.,][]{davit2013homogenization}. Other upscaling approaches are described in \citep{brenner1986transport}.

 Yet, the need for computationally efficient  \deleted{`ultralong' time} predictions of subsurface system response to unsteady, and potentially highly fluctuating, forcing factors calls for the formulation of spatiotemporally-upscaled models. The practical need of averaging in time (as well as in space) originates from the disparity in temporal scales between the frequency at which the system is driven and the temporal horizon in which predictions are made, e.g., local micro-climate (precipitation, etc.) and the temporal scale relevant for climate studies, or local human activity and CO$_2$ sequestration scenarios, \added{which, will be referred to as `long times' in the following}.  In an attempt to curb computational burden, this problem is often tackled by adopting larger time-stepping and by temporally averaging time-dependent boundary conditions or driving forces \added{\citep{Beese-1980-solute,wang-2009-effects,yin-2015-impacts}}.

 While standard in the theory of turbulence \citep{Taylor-1959-present,pope},  time-averaging of fine-scale models of flow in porous media and geologic formations  has attracted \deleted{much} less attention \deleted{compared to its overbearing spatial-averaging sibling} \citep{he-1996-spatial, pavliotis2002homogenized, rajabi2015spatio, rajabi2017frequency, rajabi2021stochastic}.  Yet, the implications of temporally unresolved boundary conditions and driving forces in nonlinear subsurface systems appear to be dire: \added{for example,} Wang \emph{et al.} \citep{wang-2009-effects}  \deleted{demonstrated} \added{showed}  that  predictions of nonlinear transport in the vadose zone are greatly affected by the time resolution of forcing factors (e.g. annual versus hourly meteorological data). In partially saturated flows, \cite{Bresler-1982-Unsaturated} and \cite{Russo-1989-Numerical} found  breakthrough curves under time-varying and time-averaged boundary conditions to be very different, with contaminant travelling faster and further in the former case. Similar highly dynamical conditions can be found in the subsurface interaction zone (SIZ) of riverine systems where environmental transitions often result in biogeochemical hotspots and moments that drive microbial activity and control organic carbon cycling \citep{Stegen-2016-Groundwater}. \added{To the best of our knowledge,} with a few exceptions \citep{Beese-1980-solute,he-1996-spatial,pavliotis2002homogenized,wang-2009-effects}, the effects of temporal averaging on macroscopic transport \deleted{have been overlooked} \added{have not been thoroughly investigated. On the contrary, the impact of temporally fluctuating flows, boundary conditions and forcings in the context of upscaled transport in porous media has been the object of a number of studies. The seminal work by \cite{Smith-1981-delay,smith-1982-contaminant} investigated the impact of dispersion in oscillatory flows and derived a spatially upscaled delay-diffusion equation which accounts for memory effects. The effect of periodic oscillations leads to dynamic effective dispersion and time-dependent closure problems as analyzed by a number of authors \citep[e.g.][]{moyne-1997-two,parada-2011-Frequency,davit-2012-comment,parada-2011-reply,dentz2003effective,pool2014effects,pool2015effects,pool2016transient,Nissan-2017-time}.}  Other studies \deleted{primarily} focused  on spatial upscaling of transport in porous media \deleted{with fluctuating flows  and} with changing pore-scale geometry due, e.g., to precipitation/dissolution processes  \citep{Noorden-2010-upscaled,Kumar-2011-effective,Kumar-2014-Upscaling,Bringedal-2016-upscaling}. In the context of atmospheric and oceanic pollutant transport where  large-scale mean flow interacts non-linearly with  small-scale fluctuations, Pavliotis \emph{et al.} \citep{pavliotis2002homogenization,pavliotis2002homogenized,pavliotis2008multiscale} use higher-order homogenization  to derive a rigorous homogenized equation and screen the temporal distribution of macroscopic quantities over long times  by selecting appropriate spatial-temporally invariant volumes of the domain over which space-time volume averaging is applied. 
 \cite{fish2004space} presents a model for wave propagation in heterogeneous media by introducing multiple space-time scales with higher order homogenization theory to resolve stability and consistency issues. 
 
 \added{Here, we are primarily interested in studying the effect of space-time averaging  on the final form of the upscaled equations for  long times (rather than early  and/or pre-asymptotyc times \citep{parada-2011-reply}), i.e. when the influence on the initial condition has been forgotten, and their corresponding regimes of validity. This knowledge is important to assess the accuracy of, e.g., numerical models wherein the temporal numerical resolution  significantly exceeds characteristic scales at which the system is driven.} \deleted{Here}\added{Specifically}, we focus  on reactive transport in undeformable porous media driven by time-varying boundary conditions, whose frequency is much larger than the characteristic time scale at which transport is studied or observed at the macroscopic scale. \deleted{A typical temporal resolution used in modern simulations significantly exceeds characteristic scales at which the system is driven.}  Some of the questions we are interested in addressing are: under which conditions (e.g. signal frequency) the instantaneous macroscopic  response of the system can be decoupled from temporally fluctuating forcing factors (e.g. temporally dependent injection rates at the boundary)? How to properly account for time-averaged boundary conditions in upscaled models? We propose to address these questions by introducing the concept of spatiotemporal upscaling in the context of asymptotic multiple scale expansions. \added{The main contribution of the paper is to explicitly address the question of whether or not, and how, space-time upscaling affects reactive transport modeling, and more importantly, if/what conditions of applicability of upscaled equations need to be satisfied for the macroscopic models to be accurate. This problem becomes of increasing importance as hydrologic modeling (and its relation to climate models) expands the time-window (from months  to years to decades and more) used for forward predictions, while the time resolution in our simulations remains constrained by computational costs.}

The manuscript is organized as follows. In Section~\ref{section:model}, we present the pore-scale model describing advective and diffusive transport of a solute subject to time-dependent Dirichlet  conditions at the macroscale boundary and undergoing a heterogenous chemical reaction with the solid matrix. In Section~\ref{sec:SThomogenization}, we introduce the concept of spatiotemporal upscaling in the context of homogenization by multiple-scale expansions, and demonstrate the impact of time-dependent forcings and boundary conditions on macroscopic reactive transport.  We first classify the macroscopic dynamics in three regimes (slowly, moderately and highly fluctuating regimes) and then derive a set of frequency-dependent conditions under which scales are separable. Section~\ref{sec:discussion} provides a physical interpretation of the key analytical results of Section~\ref{sec:SThomogenization}. In Section~\ref{sec:special_cases}, we discuss different transport regimes in terms of relevant dimensionless numbers. Conclusions and  outlook  are given in Section~\ref{sec:conclusions}.

\section{Problem Formulation}\label{section:model}
\subsection{Domain and governing equations}
Let  $\hat \Omega$ be a domain in $\mathcal{R}^n$ ($n\geq 2$), bounded by $\partial\hat\Omega$, of characteristic length $L$ such that $\hat \Omega=\hat\Omega_s\cup\hat \Omega_p$, where $\hat\Omega_s$ and $\hat\Omega_p$ are the solid and pore phases in  $\hat\Omega$, respectively, and $\hat\Omega_p$ is fully saturated with a viscous fluid.  The boundary between the solid and the pore space domains is  $\hat\Gamma$. The domain $\hat\Omega$ is composed of repeating unit cells $\hat Y=\hat{\mathcal B}\cup\hat{\mathcal G}$ of characteristic size $l$ with $l\ll L$, where $\hat{\mathcal G}$ and $\hat{\mathcal B}$ are the solid and pore phases in $\hat Y$, respectively. The geometric scaling parameter 
\begin{align}
\varepsilon:=\dfrac{l}{L}\ll 1
\end{align}
 relates the size of the pore-scale unit cell to the corresponding macroscale (or observation spatial scale). 

The laminar incompressible flow of a viscous fluid through the pore space $\hat \Omega_p$ satisfies Stokes law and the continuity equation
\begin{linenomath*}
\begin{subequations}\label{eq:stokes}
\begin{align}
&{\mu}{\hat\nabla}^2 \hat{\mathbf v}_\varepsilon-{\hat \nabla}{\hat p_\varepsilon}=0,\quad \hat{\mathbf x}\in\hat \Omega^\varepsilon_p,\\
&\hat\nabla\cdot \hat{\mathbf v}_\varepsilon=0,\quad \hat{\mathbf x}\in\hat \Omega^\varepsilon_p,
\end{align}
\end{subequations}
\end{linenomath*}
subject to
\begin{linenomath*}
\begin{align}
\hat{\mathbf v}_\varepsilon=0,\quad \mathbf x\in{\hat\Gamma}^\varepsilon,\label{eq:dimensional flow}
\end{align}
\end{linenomath*}
\added{and appropriate boundary conditions on $\mathbf {v}_\varepsilon$ and $\hat p_\varepsilon$ on the domain boundary $\partial\hat\Omega$. In \eqref{eq:stokes} and \eqref{eq:dimensional flow},}
\deleted{where} $\hat{\mathbf v}_\varepsilon$ [LT$^{-1}$], $\hat p_\varepsilon$ and $\mu$ are the fluid velocity, dynamic pressure and  dynamic viscosity, respectively.
The transport of a reactive solute  $\mathcal{M}$, dissolved in the fluid, with molar concentration $\hat c_\varepsilon(\hat{\mathbf x},\hat t)$ [\added{mol}L$^{-3}$] at  $\hat{\mathbf x}\in \hat\Omega^{\varepsilon}_p$ and time $\hat t>0$, is governed by
\begin{linenomath*}
\begin{align}\label{eq:dimensional transport}
\dfrac{\partial{\hat c}_\varepsilon}{\partial{\hat t}}+\hat{\mathbf v}_\varepsilon\cdot{\hat\nabla}\hat{c}_\varepsilon={\hat\nabla}\cdot(\hat{\textbf{D}}{\hat \nabla}\hat{c}_\varepsilon),\quad\hat{\mathbf x}\in{\hat \Omega}^\varepsilon_p, \quad\hat t>0\end{align}
\end{linenomath*}
where $\hat{\mathbf{D}}$ [L$^2$T$^{-1}$] is the molecular diffusion tensor, \added{$[\mathbf D\nabla c_\varepsilon]_i=D_{ij}\partial_{x_j} c_\varepsilon$ is a matrix-vector multiplication, and `$\cdot$' represents a scalar product, e.g. $\hat{\mathbf v}_\varepsilon\cdot{\hat\nabla}\hat{c}_\varepsilon=\hat{v}_{\varepsilon, i}\partial_{x_i} c_\varepsilon$, where summation is implied over a repeated index.} The nonlinear heterogenous precipitation/dissolution reaction of solute $\mathcal{M}$ at the solid grains boundary can be modelled through the following boundary condition  on $\Gamma$
\begin{linenomath*}
\begin{align}
-\mathbf n\cdot{\hat{\textbf{D}}}{\hat \nabla}{\hat c}_\varepsilon={\hat k}({\hat c}^a_\varepsilon-\overline{c}^a)\qquad \hat{\mathbf x}\in{\Gamma}^\varepsilon\label{eq:dimensional BC}
\end{align}
\end{linenomath*}
which represents a mass balance  across the solid-liquid interface. Equation \eqref{eq:dimensional transport} is subject to initial conditions
\begin{linenomath*}
\begin{align}
\hat{c}_\varepsilon(\hat{\mathbf x},0)=\hat c_{\mbox{\tiny{in}}}(\hat{\mathbf x}), \quad \hat{\mathbf x}\in \hat\Omega_p
\end{align}
\end{linenomath*}
and boundary conditions on $\partial \hat\Omega=\partial \hat\Omega_D\cup\partial \hat\Omega_N\cup\partial \hat\Omega_R$, where $\partial \hat\Omega_i$, $i=\{D,N,R\}$ represent a portion of the boundary subject to Dirichlet, Neumann or Robin boundary conditions, respectively.  Without loss of generality, we assume $\partial\hat\Omega_D$ is subject to time-varying boundary conditions, i.e.
\begin{linenomath*}
\begin{align}
\hat{c}_\varepsilon(\hat{\mathbf x}_D,\hat t)=\hat c_{D}(\hat t).
\end{align}
\end{linenomath*}
\deleted{where $\hat \tau$ is the characteristic time scale of the boundary forcing} \added{The previous boundary condition  models a spatially localized  seasonal release of reacting solute (e.g. contaminant or nutrient),  associated to, e.g., respiration processes of bacteria, hydrologic cycles that create local chemical hotspots (e.g. in the hyporheic corridor), etc. We emphasize that other time-dependent boundary conditions could be used in place of \eqref{bc1}, e.g. Danckwerts' \citep{Danckwerts-1953-Continuous}.}
\subsection{Dimensionless formulation}
We define the following  dimensionless quantities
\begin{linenomath*}
\begin{align}
c=\dfrac{{\hat c}_\varepsilon}{\hat c_{\mbox{\tiny{in}}}},\quad\textbf{D}=\dfrac{{\hat{\mathbf D}}}{D},\quad \mathbf x=\dfrac{{\hat{\mathbf x}}}{L},\quad \mathbf v_\varepsilon=\dfrac{\hat{\mathbf v}_\varepsilon}{U},\quad t=\dfrac{{\hat t}}{\tau_c}, \quad p=\dfrac{{\hat p}l^2}{{\nu} U L}\label{eq:nondimensionalization definitons}
\end{align}
\end{linenomath*}
where $U$, $D$ and $\tau_c$ are characteristic scales for velocity, diffusivity and time. We set $\tau_c$ as the diffusive time-scale, i.e.
\begin{linenomath*}
\begin{align}\label{eq:diff-time}
\tau_c=\dfrac{L^2}{D},
\end{align}
\end{linenomath*}
Inserting \eqref{eq:nondimensionalization definitons} and \eqref{eq:diff-time} in \eqref{eq:stokes}-\eqref{bc1}, one obtains 
\begin{linenomath*}
\begin{align}
&\varepsilon^2\nabla^2\mathbf v_\varepsilon-\nabla  p_{\varepsilon}=0 \quad \mbox{and}\quad \nabla\cdot\mathbf v_\varepsilon=0, \quad  \mathbf x\in{\Omega}^\varepsilon_p\label{eq:stokes-dimensionless}
\end{align}
\end{linenomath*}
subject to
\begin{linenomath*}
\begin{align}
&\mathbf v_\varepsilon=0,\quad \mathbf x\in\Gamma^\varepsilon,\label{eq: flow noslip condition}
\end{align}
\end{linenomath*}
and 
\begin{linenomath*}
\begin{align}
& \dfrac{\partial c_\varepsilon}{\partial t}+\nabla \cdot(-\textbf{D}\nabla c_\varepsilon+\mbox{Pe} \mathbf v_\varepsilon c_\varepsilon)=0, \quad\mathbf x\in{\Omega}^\varepsilon_p, \quad t>0\label{eq:nondimensioal transport}
\end{align}
\end{linenomath*}
subject to
\begin{linenomath*}
\begin{align}
-\mathbf n\cdot\textbf{D}\nabla c_\varepsilon&=\mbox{Da}(c^a_\varepsilon-1)\qquad \mathbf x\in\Gamma^\varepsilon,\quad t>0\label{eq:nondimensioal BC}\\
c(\mathbf x,0)& =c_{\mbox{\tiny{in}}}(\mathbf x)\qquad \mathbf x\in \Omega_p^\varepsilon,\label{eq:initial condition}
\end{align}
\end{linenomath*}
and to time-varying Dirichlet boundary conditions on a subset of the macroscopic boundary $\partial \Omega_D$, i.e.
\begin{linenomath*}
\begin{align}\label{bc1}
c_\varepsilon(\mathbf x_D,t)= c_{D}(t).
\end{align}
\end{linenomath*}
In \eqref{eq:nondimensioal transport} and \eqref{eq:nondimensioal BC}
 \begin{linenomath*}
\begin{align}\label{Pe_Da}
\mbox{Pe}:=\dfrac{\tau_d}{\tau_a}=\dfrac{U L}{D},\qquad \text{and}\qquad \mbox{Da}:=\dfrac{\tau_d}{\tau_r}=\dfrac{L \hat{k} \hat c^{a-1}_0}{D},
\end{align}
\end{linenomath*}
are the P\'{e}clet and Damkh\"{o}ler numbers, defined as the ratio between the diffusive time $\tau_d$ and the advection and reaction time scales, $\tau_a$ and $\tau_r$, respectively, with $\tau_a=L/U$ and $\tau_r=L/(\hat k \hat c_0^{a-1})$.
%
\section{Space-Time Homogenization via Multiple-Scale Expansions}\label{sec:SThomogenization}
In this section, we generalize the multiple-scale expansion method to upscale in both space and time the pore scale dimensionless equations \eqref{eq:stokes-dimensionless} and \eqref{eq:nondimensioal transport} to the macroscale, and to derive effective equations for the space-time averages of the flow velocity $\langle \mathbf v_\varepsilon \rangle$ and the solute concentration $\langle  c_\varepsilon \rangle$ while accounting for time-varying boundary conditions. We emphasize a similar approach can be employed to handle time-varying source terms and coefficients.

\subsection{Preliminaries and Extensions to Time Homogenization}\label{preliminaries}
Within the multiple-scale expansion framework, we introduce a `fast' space variable $\mathbf y$ defined in the unit cell $Y$, i.e. $\mathbf y\in Y$. Furthermore, if the system is driven by time-varying boundary conditions or forcing factors with characteristic time scale $\hat \tau\ll  T$  where $T$ is the observation time scale, one can   define a temporal scaling parameter
\begin{linenomath*}
\begin{align}
\omega:=\dfrac{\hat \tau}{T}\ll 1,
\end{align}
\end{linenomath*}
that relates the driving force/boundary condition frequency ($\sim 1/\hat \tau$) and the observation (macroscopic) time scale $T$.  We define the exponent $\gamma$ such that
\begin{linenomath*}
\begin{align}\label{gamma-epsilon}
\varepsilon=\omega^\gamma,
\end{align}
\end{linenomath*}
i.e. $\gamma$ quantifies the separation between temporal and spatial scales and is uniquely determined once the characteristic length and time scales of the problem are identified. Each variable is defined as follows,
\begin{linenomath*}
\begin{align}
\mathbf y = \varepsilon^{-1}\mathbf x , \quad \mbox{and} \quad \tau  = \omega^{-1} t.
\end{align}
\end{linenomath*}
For any pore-scale quantity $\psi_\varepsilon$,
\begin{linenomath*}
\begin{align}\label{space-averages}
 \langle \psi_\varepsilon \rangle_Y \equiv \frac{1}{|Y|} \int\limits_{\mathcal{B}(\mathbf x)}\psi_\varepsilon \mathrm{d}\mathbf{y}, \quad
 \langle \psi_\varepsilon \rangle_\mathcal{B} \equiv \frac{1}{|\mathcal{B}|} \int\limits_{\mathcal{B}(\mathbf x)} \psi_\varepsilon  \mathrm{d}\mathbf{y},
\, \mbox{and}\,\,
\langle \psi_\varepsilon \rangle_\Gamma \equiv \frac{1}{|\Gamma|} \int\limits_{\Gamma(\mathbf x)} \psi_\varepsilon \mathrm{d}\mathbf{y}
\end{align}
\end{linenomath*}
are three local spatial averages (function of $\mathbf x$) over the pore space $\mathcal B(\textbf x)$ of the unit cell $Y(\textbf x)$ centered at $\textbf x$. In \eqref{space-averages}, $\langle \psi_\varepsilon \rangle_Y= \phi \langle\psi_\varepsilon \rangle_\mathcal{B}$ and $\phi=|\mathcal{B}|/|Y|$ is the porosity. Similarly, one can define temporal averages (function of $t$) over a time unit cell $\mathcal I$ centered at $t$, i.e,
 \begin{linenomath*}
 \begin{align}\label{time-average}
 \langle \psi_\varepsilon \rangle_{\mathcal I} \equiv \frac{1}{|\mathcal I|} \int\limits_{\mathcal{I}(t)}\psi_\varepsilon \mathrm{d}\tau.
\end{align}
\end{linenomath*}
where $\mathcal I$ is the smallest time-scale resolved at the macroscale, e.g. the discretization time-step at the continuum scale.
The space-time averages $\langle \psi_\varepsilon \rangle_{\mathcal {IB}}$ \added{and $\langle \psi_\varepsilon \rangle$ are} \deleted{is} defined as
\begin{linenomath*}
\begin{align}
\langle \psi_\varepsilon \rangle_{\mathcal {IB}}:=\langle \langle \psi_\varepsilon \rangle_{\mathcal I} \rangle_{\mathcal B}=\langle \langle \psi_\varepsilon \rangle_{\mathcal B} \rangle_{\mathcal I}.
\end{align}
\end{linenomath*}
\added{and}
\begin{linenomath*}
\begin{align}
\langle \psi_\varepsilon \rangle:=\langle \langle \psi_\varepsilon \rangle_{\mathcal I} \rangle_Y=\langle \langle \psi_\varepsilon \rangle_Y \rangle_{\mathcal I}=\phi \langle \psi_\varepsilon \rangle_{\mathcal {IB}}.
\end{align}
\end{linenomath*}
Furthermore, any pore-scale function $\psi_\varepsilon(\mathbf{x}, t)$ can be represented as  $\psi_\omega(\mathbf{x}, t)$ through \eqref{gamma-epsilon} and $\psi_\omega\left(\mathbf{x},t \right) :=\psi(\mathbf{x},\mathbf{y},t,\tau)$. Replacing $\psi_\omega\left(\mathbf{x},t \right)$ with $\psi(\mathbf{x},\mathbf{y},t,\tau_\mathrm{a},\boldsymbol \tau_\mathrm r)$ gives the following relations for the spatial and temporal derivatives,
\begin{linenomath*}
\begin{align}\label{eq:dx}
\nabla \psi_\omega = \nabla_\mathbf x \psi + \varepsilon^{-1}\nabla_\mathbf y \psi= \nabla_\mathbf x \psi + \omega^{-\gamma}\nabla_\mathbf y \psi, \quad \mbox{and} \quad \dfrac{\partial \psi_\omega}{\partial t}=\dfrac{\partial \psi}{\partial t}+\omega^{-1}\dfrac{\partial \psi}{\partial \tau}
\end{align}
\end{linenomath*}
respectively. The function $\psi$ is represented as an asymptotic series in powers of $\omega$,
\begin{linenomath*}
\begin{align}\label{asymp_exp}
\psi(\mathbf{x},\mathbf{y},t, \tau) = \sum_{m=0}^\infty \omega^m \psi_m(\mathbf x, \mathbf y, t, \tau),
\end{align}
\end{linenomath*}
wherein $\psi_m(\mathbf x, \mathbf y, t, \tau)$, $m=0,1,\ldots$, are $Y$-periodic in $\mathbf y$. Finally, we set
\begin{linenomath*}
\begin{align}\label{da-eps}
\mbox{Pe} = \omega^{-\alpha},\quad\mbox{and}\quad
\mbox{Da} = \omega^\beta, 
\end{align}
\end{linenomath*}
with the exponents $\alpha$ and $\beta$ determining the system behavior. We seek the asymptotic space-time average behavior of $\psi_\omega$ as $\omega\rightarrow 0$ for any arbitrary time-scale separation parameter $\gamma$.\\
 It should be emphasized that, an important step in solving the cascade of equations for $\psi_0, \psi_1, \cdot\cdot\cdot$, is consistently checking whether the solvability condition is satisfied. Otherwise, the derivation leads to misleading results. More specifically, when seeking a solution for $\psi_1$, we have to impose the solvability condition. This condition ensures existence and uniqueness of a solution rigorously by employing the \textit{Fredholm Alternative}. Critical points to consider while employing the homogenization theory to upscale the transport equation are summarized by \cite{auriault2019comments}, where the author explicitly mentions that the averaging process is imposed by Fredholm Alternative, and there is no arbitrary step along the derivation process.
%
\subsection{Upscaled Transport Equations and Homogenizability Conditions}\label{Sec:Homogenizability Conditions}

The homogenization of the Stokes equations \eqref{eq:stokes} leads to the classical result 
%
\begin{linenomath*}
	\begin{align}
		\langle \mathbf v\rangle=-\mathbf{K}\cdot\nabla P_0,\qquad\nabla\cdot\langle \mathbf v\rangle=0, \quad \mathbf x\in \Omega,\label{eq:upscaled flow}
	\end{align}
\end{linenomath*}
where the dimensionless permeability tensor $\mathbf K$ is defined as $\mathbf{K}=\langle \mathbf k\rangle $ and $\mathbf k$ is the closure variable, solution of the closure problem 
\begin{linenomath*}
	\begin{align}
\nabla^2\textbf k+\mathbf I-\nabla \textbf a=0,\qquad\nabla\cdot\mathbf k=0,\qquad \textbf y\in \mathcal B
	\end{align}
\end{linenomath*}
subject to $\mathbf k(\textbf y)=0$  for $\textbf y\in \Gamma$ and $\langle \textbf a\rangle=0$, where $\mathbf a$ is $Y$-periodic \cite[pp. 46-47, Theorem 1.1]{hornung2012homogenization}.

\added{Here, we are interested in studying the system for long times, also referred to as `quasi-steady stage' (as per definition of \cite{parada-2011-reply}), i.e. when  both  time- and length-scales can be separated. Then}\deleted{as detailed in Appendix \ref{AppendixA}}, the space-time homogenization of the pore-scale reactive transport equations \eqref{eq:nondimensioal transport}-\eqref{bc1} up to order $\omega^2$, leads to \citep{rajabi2021stochastic} \added{(details in Appendix \ref{AppendixA})}
\begin{linenomath*}
	\begin{align}
		\phi\dfrac{\partial \langle c\rangle_{\mathcal{IB}}}{\partial t}=\nabla\cdot\left[\tilde{\tilde{\textbf{D}}}^\star\nabla\langle c\rangle_{\mathcal {IB}}-\mbox{Pe}\langle c\rangle_{\mathcal{IB}}\langle\mathbf v\rangle_{\mathcal{IB}}\right]+\phi \omega^{-\gamma}\mathcal{K}^{\star}\mbox{Da}(1-\langle c\rangle_{\mathcal{IB}}^a),\nonumber \\  (\mathbf x,t)\in \Omega\times(0,T), \label{eq:upscaled eq}
			\end{align}
\end{linenomath*}
where the effective coefficients $\mathcal{K^\star}$, and  $\tilde{\tilde{\textbf{D}}}^\star$ are defined as%
\begin{linenomath*}
	\begin{align}
		\mathcal{K^\star}&=\dfrac{|\varGamma|}{|\mathcal B|},\\
		\tilde{\tilde{\textbf{D}}}^\star&=\langle\textbf{D}(\mathbf I+\omega^{1-\gamma}\nabla_{\mathbf y}\boldsymbol{\chi})\rangle+\omega^{1-\alpha}\langle\boldsymbol{\chi}\mathbf k\rangle\cdot\nabla_{\mathbf x}P_0,\label{eq:dispersion D star}
	\end{align}
\end{linenomath*}
and $\boldsymbol{\chi}(\mathbf y,\tau)$ \deleted{and $\lambda(\mathbf y,\tau)$ are} is the closure variable. The effective coefficient $\tilde{\textbf{D}}^\star$ \deleted{and $\tilde{\textbf{D}}'$ are} is computed through the solution of \deleted{two} \added{the unsteady} auxiliary  cell problem for \deleted{$\lambda(\mathbf y,\tau)$ and}  $\boldsymbol{\chi}(\mathbf y,\tau)$, i.e.
\begin{linenomath*}
	\begin{align}\label{eq:closure probs and BCS}
&		\dfrac{\partial\boldsymbol{\chi}}{\partial \tau}+\omega^{-\alpha}(\mathbf v_0-\langle\mathbf v_0\rangle_{\mathcal I\mathcal B})-\omega^{-\gamma}\nabla_{\mathbf y}\cdot\textbf{D}(\mathbf I+\omega^{1-\gamma}\nabla_{\mathbf y}\boldsymbol{\chi})+\omega^{1-\gamma-\alpha}\mathbf v_0\cdot(\nabla_{\mathbf y}\boldsymbol{\chi})=0, \quad \mathbf y\in \mathcal B,\nonumber\\
&		\mathbf n\cdot\textbf{D} (\mathbf I+\omega^{1-\gamma}\nabla_{\mathbf y}\boldsymbol{\chi})=0, \quad \mathbf y\in \mathcal B,\\
&		\boldsymbol\chi(\mathbf y, 0)=\boldsymbol\chi_{\mbox{\tiny{in}}}(\mathbf y),\nonumber
	\end{align}
\end{linenomath*}
and  $\langle\boldsymbol{\chi}\rangle_{\mathcal B}=0$,  where $\mathbf v_0=-\mathbf k(\mathbf y)\cdot\nabla_{\mathbf x} P_0$ is the solution of the homogenized flow equation \eqref{eq:upscaled flow}, provided the following conditions are met \citep{rajabi2021stochastic}

\begin{enumerate}
	\item $\varepsilon\ll 1$,
	\item $\omega\ll 1$,
	\item $\langle \boldsymbol{\chi}\rangle_{\mathcal{I}\Gamma}\approx\langle\boldsymbol{\chi}\rangle_{\mathcal{IB}}$,
	\deleted{ \item $\langle \lambda\rangle_\Gamma\approx\langle\lambda\rangle_{\mathcal B}$,}
\end{enumerate}
\noindent Additional bounds on the Damk\"{o}hler and P\'{e}clet numbers must be satisfied depending on the time-space scale separation parameter $\gamma$. Specifically,
\begin{enumerate}
\item[5a.] when $\varepsilon<\omega$, i.e. $\gamma>1$, the system is referred to as \emph{slowly fluctuating} and the additional conditions  to guarantee that scale separation occurs are 
\begin{enumerate}
\item $\mbox{Pe}<\omega^{-1}$
\item $\mbox{Da}/\mbox{Pe}<\varepsilon$
\item $\mbox{Da}<\varepsilon$.
\end{enumerate}
\item[5b.] When $\omega <\varepsilon<\omega^{1/2}$ (or $\omega\approx\varepsilon$), i.e. $1/2<\gamma<1$, the system is referred to as \emph{moderately fluctuating} and the additional conditions  to guarantee that scale separation occurs are 
\begin{enumerate}
\item $\dfrac{\varepsilon}{\omega}<\mbox{Pe}<\omega^{-1}$
\item $\mbox{Da}/\mbox{Pe}<\omega$.
\end{enumerate}
\item[5c.] When $\omega^{1/2}<\varepsilon<1$ (or $\varepsilon\gg\omega$), i.e. $0<\gamma<1/2$, the system is referred to as \emph{highly fluctuating} and the additional conditions  to guarantee that scale separation occurs are 
\begin{enumerate}
\item $\mbox{Pe}<\omega^{-1}$
\item $\mbox{Da}/\mbox{Pe}<\omega$
\item $\mbox{Da}<\omega/\varepsilon$.
\end{enumerate}
\end{enumerate}

\begin{figure}[!ht]
  	\centering
  	\includegraphics[width=\textwidth]{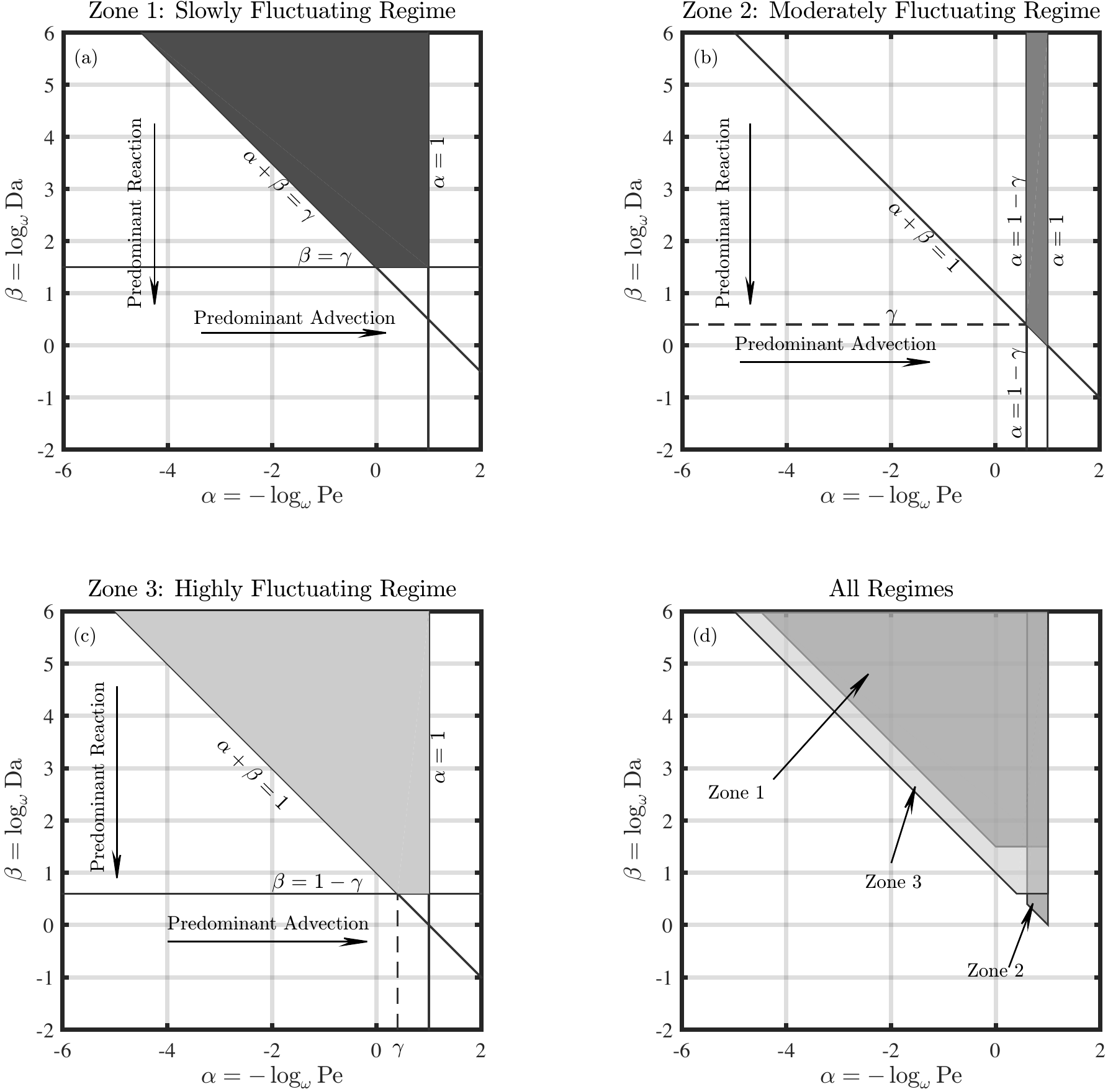}
	 	\caption{Applicability conditions in the $(\alpha,\beta)$-phase space for: (a) \emph{slowly fluctuating regimes} (Zone 1), i.e. $\varepsilon<\omega$ or $\gamma>1$; (b) \emph{moderately fluctuating regimes} (Zone 2), i.e. $\omega<\varepsilon<\omega^{1/2}$ ($\omega\approx\varepsilon$) or $1/2<\gamma<1$; (c) \emph{highly fluctuating regimes} (Zone 3), i.e. $\omega^{1/2}<\varepsilon<1$ ($\varepsilon\gg \omega$) or $0<\gamma<1/2$; (d) all regimes overlapped for direct comparison. In each Figure, the shaded region identifies sufficient conditions for the validity of macroscopic equation in terms of Da and Pe numbers. In the white region, scales are not well separated and macroscopic and microscopic models should be solved simultaneously.}\label{All_New}
  \end{figure}

These conditions can be graphically visualized in a phase diagram in the $(\mathrm{Pe},\mathrm{Da})$-space, or the $(\alpha,\beta)$-space for the three different regimes \citep{rajabi2021stochastic}. The bounds for  slowly fluctuating  systems (i.e. $\varepsilon<\omega$, i.e. $\gamma>1$) are summarized in the $(\alpha,\beta)$-plane of Figure~\ref{All_New}(a), where the lines $\beta=\gamma$, $\alpha+\beta=\gamma$ and  $\alpha=1$     correspond to $\mbox{Da}=\varepsilon$, $\mbox{Da}/\mbox{Pe}=\varepsilon$  and $\mbox{Pe}=\omega^{-1}$, respectively. For moderately fluctuating systems where $\omega\approx \varepsilon$, the bounds are summarized  in the $(\alpha,\beta)$-plane of Figure~\ref{All_New}(b). The lines  $\alpha+\beta=1$, $\alpha=1$ and $\alpha=1-\gamma$ correspond to $\mbox{Da}/\mbox{Pe}=\omega$, $\mbox{Pe}=\omega^{-1}$ and $\mbox{Pe}=\varepsilon/\omega$, respectively. Finally, in highly fluctuating systems, i.e. when $\omega^{1/2}<\varepsilon<1$ (or $\varepsilon\gg\omega$), the previous conditions are summarized in Figure~\ref{All_New}(c), where the line $\beta=1-\gamma$ corresponds to $\mbox{Da}=\omega/\varepsilon$. \added{Figure~\ref{All_New}(d) overlaps the applicability conditions for the three regimes to allow direct comparison.}
\added{We emphasize that, while Eq.~\eqref{eq:upscaled eq} has the  form of a classical advection-reaction-dispersion equation, both the (i) the form of its effective coefficients  and (ii)  the conditions under which both spatial and temporal scales are fully decoupled explicitly depend on $\gamma$, i.e. the scale parameter that relates spatial scales and the frequency of the boundary fluctuations. Furthermore, Eq.~\eqref{eq:upscaled eq} is consistent with the results obtained  through space and time volume averaging (ST-averaging) by \cite{he-1996-spatial} where, for a homogeneous distribution of elementary (space-time averaging) domains, the  ST-upscaling using volume averaging degenerates into a classical volume average.}

 \section{Discussion and Physical Interpretation}\label{sec:discussion}
\begin{figure}[h!]
		\includegraphics[width=\textwidth]{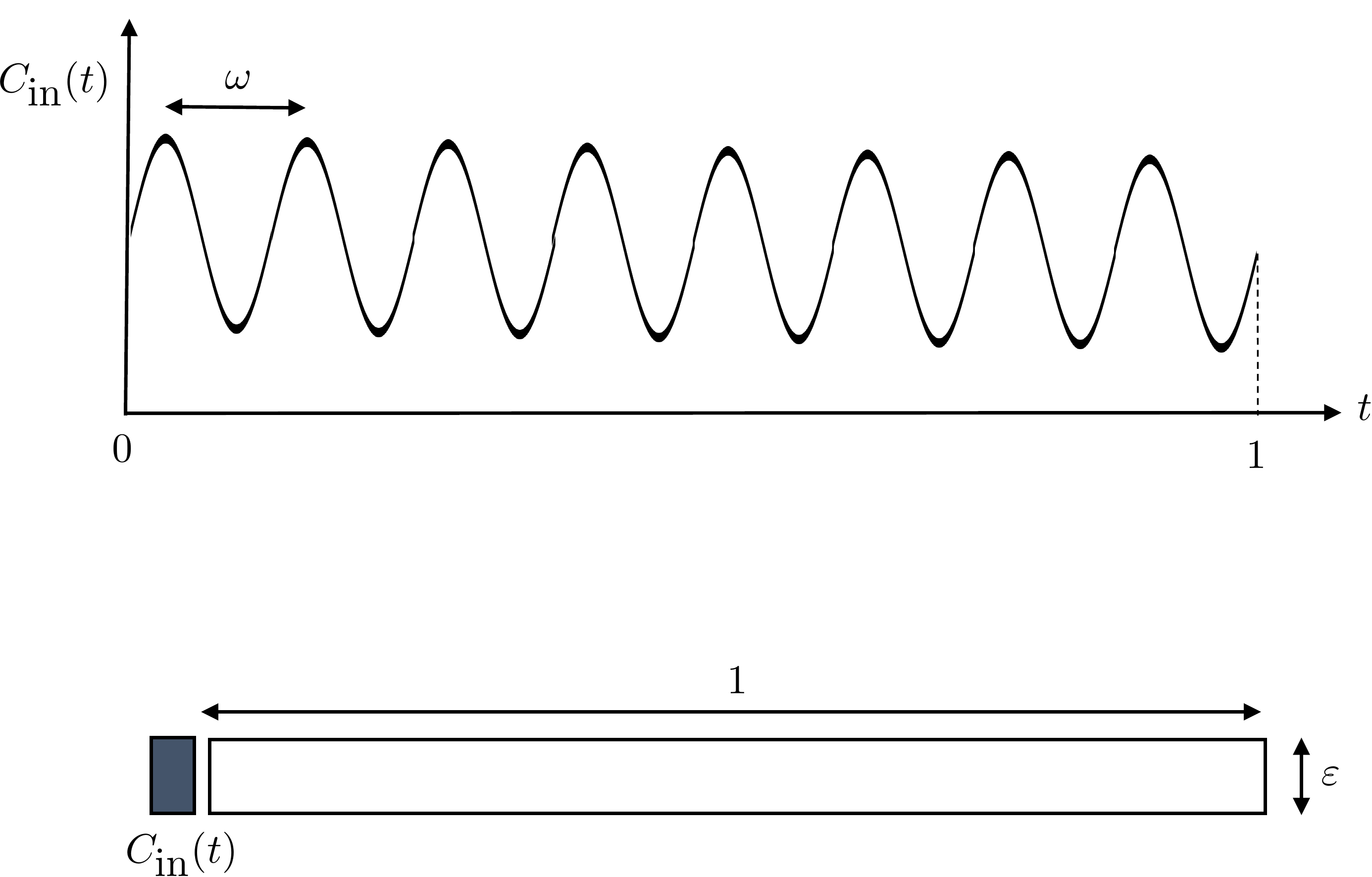}
		\caption{Time-dependent solute injection boundary condition $(C_{\mbox{in}})$ (top) at the inlet of a planar thin impermeable fracture of aperture $\varepsilon\ll1$ (bottom). The time-varying boundary condition has a frequency of $\omega^{-1}$ (or characteristic dimensionless time scale/period $\omega\ll1$). Figure not in scale.}\label{frequency-geometry}
\end{figure}

  \added{In this Section, we are concerned with providing a physical interpretation of the (formally derived) thresholds on $\gamma$ and their connection with the regimes classification (slowly, moderately and highly fluctuating regimes) proposed in the previous Section. For this purpose, we consider a conceptual example, which, despite its simplicity, maintains enough complexity to provide useful physical insights on the theoretical results.}
Without loss of generality, let us consider a pressure-driven flow through a thin bidimensional channel of length $L$ and aperture $l$ with $l\ll L$. For a channel of width $l$, the length $L$ is to be interpreted as the ``observation scale''. Steady state fully-developed incompressible flow is assumed. Reactive solute transport at the pore-scale is governed by \eqref{eq:dimensional transport} subject to \eqref{eq:dimensional BC} on the fracture walls. Time varying Dirichlet boundary conditions for solute concentration are imposed at the fracture inlet. The characteristic time scale of the fluctuating boundary conditions is $\hat \tau\ll T$, with $T$ the macroscale observation time. Figure~\ref{frequency-geometry} shows a sketch  of the system. 
As discussed in Section~\ref{preliminaries}, the space and time scale separation parameters are
\begin{linenomath*}
\begin{align}\label{time-grande}
\varepsilon\equiv{\dfrac{l}{L}}\ll1,\quad\mbox{and} \quad \omega\equiv\dfrac{{\hat \tau}}{T}\ll1.
\end{align}
\end{linenomath*}
The (dimensional)  time scales for diffusive and advective  transport at the macro-and micro-scale are
\begin{linenomath*}
\begin{subequations}
\begin{align}\label{eq:time_scales}
&\hat{t}_{\tiny{\mbox{d,macro}}}=\dfrac{L^2}{D}, \quad \hat{t}_{\tiny{\mbox{a,macro}}}=\dfrac{L}{U}, \\
&\hat{t}_{\tiny{\mbox{d,micro}}}=\dfrac{l^2}{D}, \quad \hat{t}_{\tiny{\mbox{a,micro}}}=\dfrac{l}{U},
\end{align}
\end{subequations}
\end{linenomath*}
respectively, and the (macroscopic) Pecl\'{e}t  number is defined as in \eqref{Pe_Da}
\begin{linenomath*}
\begin{align}
&\mbox{Pe}:=\dfrac{\hat{t}_{\tiny{\mbox{d,macro}}}}{\hat{t}_{\tiny{\mbox{a,macro}}}}=\dfrac{1}{\varepsilon}\dfrac{\hat{t}_{\tiny{\mbox{d,micro}}}}{\hat{t}_{\tiny{\mbox{a,micro}}}}
=\dfrac{L U}{D}.\end{align}
\end{linenomath*}
Using a diffusive scaling, i.e. $t:=\dfrac{\hat{t}}{\hat{t}_{\tiny{\mbox{d,macro}}}}$, the time scales defined in \eqref{eq:time_scales} can be expressed in terms of powers of $\varepsilon$ or $\omega$
\begin{linenomath*}
\begin{subequations}
\begin{align}\label{eq:time_scales_rescaled}
{t}_{\tiny{\mbox{d,macro}}}=\omega^0=\varepsilon^0, \qquad&  {t}_{\tiny{\mbox{d,micro}}}=\omega^{2\gamma}=\varepsilon^2, \\
 {t}_{\tiny{\mbox{a,macro}}}=\omega^\alpha=\varepsilon^{\alpha/\gamma},\qquad & {t}_{\tiny{\mbox{a,micro}}}=\omega^{\alpha+\gamma}=\varepsilon^{1+\alpha/\gamma},
\end{align}
\end{subequations}
\end{linenomath*}
\begin{linenomath*}
\begin{table}
\centering
\begin{tabular}{l|c|c}
\hline
Time scale & $\mathcal O(\omega)$ & $\mathcal O(\varepsilon)$ \\
\hline
BCs & $\omega^1$ & $\varepsilon^1$\\
${t}_{\tiny{\mbox{d,macro}}}$  &$\omega^0$ & $\varepsilon^0$ \\
 ${t}_{\tiny{\mbox{a,macro}}}$ &$\omega^\alpha$ & $\varepsilon^{\alpha/\gamma}$ \\
${t}_{\tiny{\mbox{d,micro}}}$  & $\omega^{2\gamma}$  & $\varepsilon^2$ \\
 ${t}_{\tiny{\mbox{a,micro}}}$ & $\omega^{\alpha+\gamma}$ & $\varepsilon^{1+\alpha/\gamma}$
\end{tabular}
\caption{Summary of the characteristic time scales of transport processes at the micro- and macro-scale in terms of either integer powers of $\omega$ and $\varepsilon$.}\label{time_scale}
\end{table}
\end{linenomath*}
(summarized in Table~\ref{time_scale}) and their relative magnitude is controlled by the exponents $\gamma$, $\alpha$ and $\beta$. Importantly,  the characteristic diffusion time  ${t}_{\tiny{\mbox{d,micro}}}$ scales as $\varepsilon^2$, i.e. the separation of scale parameter $\varepsilon$ can be related to the characteristic dimensionless time scale of the dominant mass transport mechanisms at the microscale. \added{This observation allows us (i) to relate the dimensionless period of the oscillations $\omega$ to the dimensionless time-scale of mass transport processes at the pore scale (specifically, diffusion), and (ii) to elucidate the physical meaning of the $\gamma$-thresholds  (i.e. $\gamma=1/2$ and $\gamma=1$) that identify the slowly, moderately and highly fluctuating regimes. Specifically,} a \emph{slowly fluctuating regime} corresponds to a system driven by time-dependent boundary conditions with a characteristic time-scale $\omega$ greater than $\varepsilon$, i.e. $\omega\gg {t}_{\tiny{\mbox{d,micro}}}$: in this regime, temporal fluctuations in the boundary conditions are very slow compared to pore-scale diffusion, and the dynamics at the microscale is exclusively controlled by local pore-scale mass transport processes. This translates in a steady state diffusive problem  for the closure variables as discussed in Section~\ref{slow}. In the \emph{moderately fluctuating regime},  $\omega< \varepsilon< \omega^{1/2}$ or, equivalently, $\omega^2< {t}_{\tiny{\mbox{d,micro}}}< \omega$, i.e.   $\omega$ and  ${t}_{\tiny{\mbox{d,micro}}}$ are of the same order of magnitude. While the local cell problems for the closure variables are still steady state (Section~\ref{moderately}), advection and diffusion become the two  mechanisms that guarantee  mixing at the pore-scale. In the \emph{highly fluctuating regime}, $\omega^{1/2}< \varepsilon< 1$ or   $\omega\ll {t}_{\tiny{\mbox{d,micro}}}$, i.e. the characteristic time scale at which the system is driven is much smaller than pore-scale diffusion time. In this regime spatial and temporal scales can still be separated, but \deleted{the high frequency boundary conditions augment local mixing and} the local cell problem becomes unsteady and advective and unsteady effects will control mass transport at the pore-scale (Section \ref{highly}). \added{It is worth noticing that the applicability domain in the (Da-Pe) space for the moderately fluctuating regimes is much smaller than those for both slowly and highly fluctuating case:  contrary to intuition, a slower advection drags the system outside the homogenizability conditions in a moderately fluctuating regime. This can be explained as follows: when diffusion and advection are the dominant mechanisms controlling transport at the pore-scale, slower advection results in an increased longitudinal, rather than transversal, mixing, making the applicability conditions in terms of Pe number much more stringent. Surprisingly, the applicability conditions in the slowly fluctuating case are a subset of those for the highly fluctuating scenario, i.e. the latter has less stringent constraints in terms of both Pe and Da numbers for the same value of $\gamma$: we hypothesize that advection and unsteadiness (and their combination) may prove more effective in achieving  pore-scale mixing, i.e. may contribute to an enhancement of mixing at the pore-scale. In presence of very fast fluctuations (at a time scale much smaller than diffusion), the pore-scale concentration in the fracture can be idealized as a periodic sequence of very thin strips of fixed concentration which travel downstream due to advection. As a result, while the system is very heterogeneous in the longitudinal direction (for times smaller than the characteristic diffusion time), it is well-mixed in the transverse direction, i.e. along the unit cell. This hypothesis is subject of current numerical investigations.}

\added{Importantly, according to \eqref{time-grande}, once the physical domain of interest is identified (i.e. $\varepsilon$ is fixed) and the characteristic time scale $\hat \tau$ of the boundary conditions determined, the macroscopic time horizon $T$ (i.e. the time at which predictions are ought to be made) uniquely defines $\omega$, and consequently, $\gamma$. This implies that, for a given pair (Pe, Da), the accuracy of the upscaled equation used for forward predictions can be greatly affected by  modifying $T$: for example if $T_2>T_1$, then $\omega_2<\omega_1\ll1$; for the same $\varepsilon$, this corresponds to $\gamma_2<\gamma_1$ since $\gamma:=\log\varepsilon/\log\omega$, i.e. the applicability conditions may change from a  slowly  to a moderately fluctuating regime. This observation suggests that caution should be employed when  systems driven by time-varying boundary conditions/forcings are \emph{de facto}, if not voluntarily, upscaled both in space and time, e.g. due to computational limitations.}

\added{In the following Section we quantitatively characterize the dominant transport mechanisms at the pore-and continuum-scale for different values of Da and Pe numbers.}
\begin{figure}[!ht]
	\centering
	\includegraphics[width=1\textwidth]{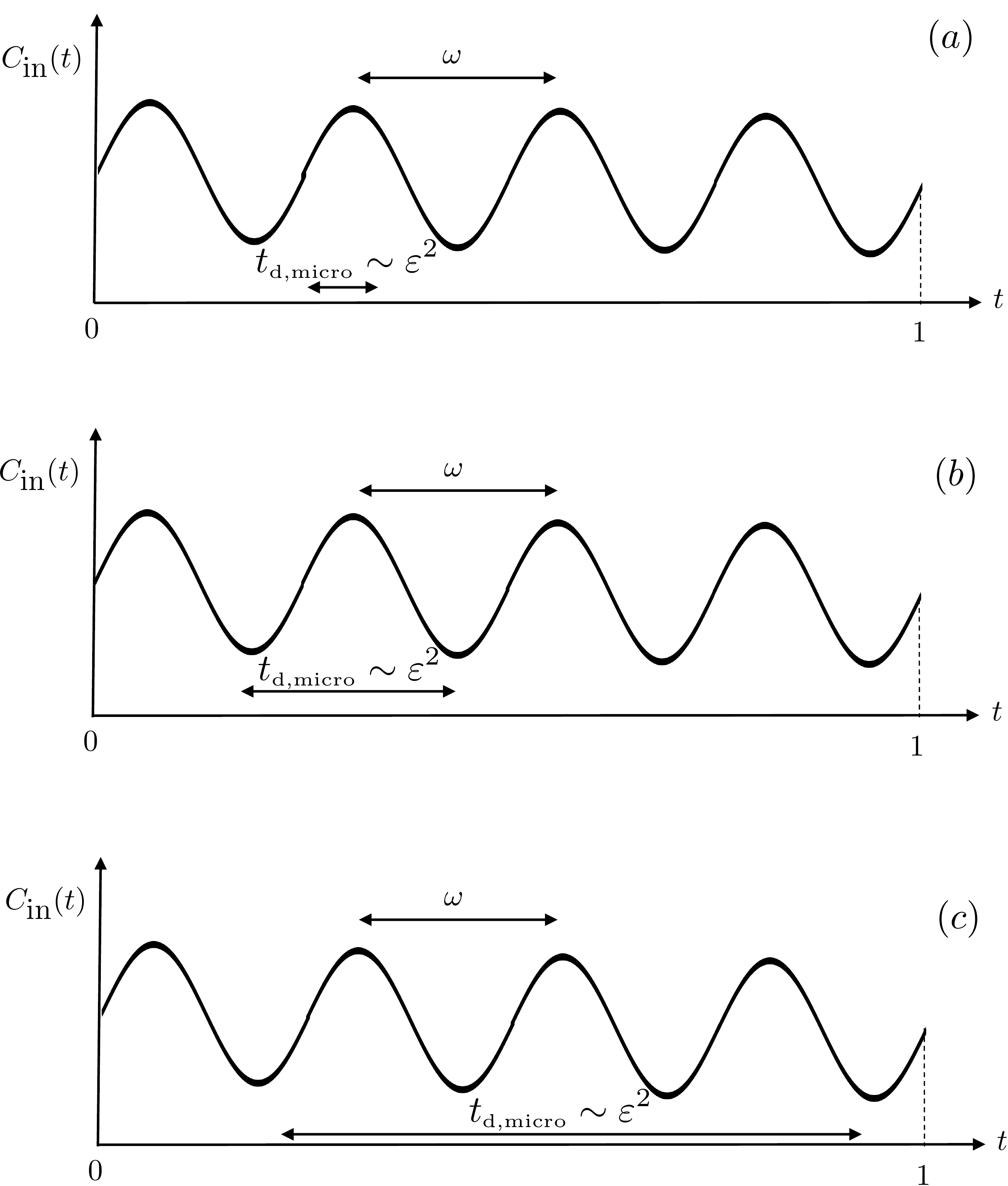}
	\caption{(a) Slowly fluctuating regime: the time-varying boundary condition $C_{\tiny{\mbox{in}}}(t)$ has a characteristic time scale much larger than pore-scale diffusion, i.e. $\omega\gg{t}_{\tiny{\mbox{d,micro}}}$. (b) Moderately fluctuating regime: the characteristic time scale of the boundary condition $C_{\tiny{\mbox{in}}}(t)$ is of the same order of pore-scale diffusion, i.e.  $\omega\approx {t}_{\tiny{\mbox{d,micro}}}$. (c)  Highly fluctuating regime: pore-scale diffusion  is much slower than the time scale imposed by $C_{\tiny{\mbox{in}}}(t)$. Figure not in scale.}
\end{figure}

\section{Special Cases}\label{sec:special_cases}
\added{In this Section, we investigate specific flow and transport regimes under which the  upscaled equation \eqref{eq:upscaled eq} and the closure problem \eqref{eq:closure probs and BCS} can be simplified. Such transport regimes are identified by the order of magnitude of the Damk\"{o}hler and Pecl\'{e}t numbers. Differently from similar analyses on the applicability conditions of diffusive-advective-reactive systems under steady boundary conditions (or forcings) \citep{auriault1995taylor} and/or the dynamics of composite materials \citep{Auriault-1991-heteronenous}, here we are specifically interested in elucidating the impact of boundary/forcing frequency on the form of the upscaled equations for the highly, moderately and slowly fluctuating regimes and  any given pair of Damk\"{o}hler and Pecl\'{e}t numbers satisfying the conditions outlined in Section~\ref{Sec:Homogenizability Conditions}. Our analysis below shows that for systems with the same Damk\"{o}hler and Pecl\'{e}t numbers, the form of the space-time upscaled equations and of the closure problem depends on the frequency of the boundary condition, i.e. pore-scale mixing is controlled by the interplay of diffusion, advection and unsteady effects (due to boundary frequency), and not by the characteristic time scales of diffusive, advective and reactive transport processes only.}
\deleted{
In this section we examine how the conditions of well-mixing at the pore level can lead to a simplification of the upscaled equations and related closure problems. For this purpose, we would investigate how the continuum scale behaves for different ranges of $\gamma$ which shows different oscillations of boundary conditions versus the spatial scale separation. As stated in the conditions, reaction is negligible at the pore level.}
\subsection{Slowly Fluctuating Boundary Conditions: $\varepsilon<\omega$}\label{slow}
\subsubsection{Transport regime with $\mbox{Pe}<1$}
In this case, Eq.\eqref{eq:upscaled eq} simplifies to a dispersion-reaction equation, since diffusion dominates advection at the macro-scale.
\begin{linenomath*}
	\begin{align}
		\phi\dfrac{\partial \langle c\rangle_{\mathcal{IB}}}{\partial t}=\nabla\cdot\left[\tilde{\tilde{\textbf{D}}}^\star\nabla\langle c\rangle_{I\mathcal B}\right]+\phi \omega^{-\gamma}\mathcal{K}^{\star}\mbox{Da}(1-\langle c\rangle_{\mathcal{IB}}^a),\label{eq:slowly negative alpha}
			\end{align}
\end{linenomath*}
where  $\tilde{\tilde{\textbf{D}}}^\star=\langle\textbf{D}(\mathbf I+\omega^{1-\gamma}\nabla_{\mathbf y}\boldsymbol{\chi})\rangle$ \deleted{and vector $\tilde{\tilde{\textbf{D}}}'=\omega^{1-\gamma}\langle\textbf{D}\nabla_{\mathbf y}\lambda\rangle$ are} \added{is} determined from the simplified closure problem
\begin{linenomath*}
\begin{subequations}\label{eq:closure-chi-diffusive}
	\begin{align}
&	\nabla_{\mathbf y}\cdot\textbf{D}(\mathbf I+\omega^{1-\gamma}\nabla_{\mathbf y}\boldsymbol{\chi})=0, \quad \mathbf y\in \mathcal B,\\
&		\mathbf n\cdot\textbf{D} (\mathbf I+\omega^{1-\gamma}\nabla_{\mathbf y}\boldsymbol{\chi})=0, \quad \mathbf y\in \mathcal B,
	\end{align}
	\end{subequations}
\end{linenomath*} 
where the advective and unsteady terms at the pore-scale can be neglected compared to the diffusive ones. In this regime, the  characteristic time scale of boundary fluctuations  is much larger than the diffusive time scale, and the system dynamics at the pore-scale is entirely controlled by diffusion processes, \added{as mentioned in Section~\ref{sec:discussion}}. This results in a steady-state purely diffusive closure problem. \added{The magnitude of the Damk\"{o}hler number Da determines the effects of chemical reactions on transport at the macroscale.}

\noindent \paragraph{Diffusion dominates reactions} $\mbox{Da}<\omega$. In this regime, diffusion dominates advection and reactive transport processes at the macro-scale as well. As result, the macroscale equation \eqref{eq:slowly negative alpha} reduces to
\begin{linenomath*}
	\begin{align}
		\phi\dfrac{\partial \langle c\rangle_{\mathcal{IB}}}{\partial t}=\nabla\cdot\left[\tilde{\tilde{\textbf{D}}}^\star\nabla\langle c\rangle_{I\mathcal B}\right].
			\end{align}
\end{linenomath*}
where the closure variable \deleted{$\lambda$ and} $\boldsymbol\chi$ still satisfies \eqref{eq:closure-chi-diffusive}.

\subsubsection{Transport regime with $1\leq\mbox{Pe}<\omega^{-1}$}
In this regime, dispersion and advection are comparable at the macroscale, and the upscaled transport equation   is~\eqref{eq:upscaled eq} with effective coefficient $\tilde{\tilde{\textbf{D}}}^\star$ \deleted{and  $\tilde{\tilde{\textbf{D}}}'$} defined by~\eqref{eq:dispersion D star} \deleted{and~\eqref{eq:dispersion D prime}}, i.e., $\tilde{\tilde{\textbf{D}}}^\star=\langle\textbf{D}(\mathbf I+\omega^{1-\gamma}\nabla_{\mathbf y}\boldsymbol{\chi})\rangle+\omega^{1-\alpha}\langle\boldsymbol{\chi}\mathbf k\rangle\cdot\nabla_{\mathbf x}P_0$ \deleted{ and $\tilde{\tilde{\textbf{D}}}'=\omega^{1-\gamma}\langle\textbf{D}\nabla_{\mathbf y}\lambda\rangle+\omega^{1-\alpha}\langle\lambda\mathbf k\rangle\cdot\nabla_{\mathbf x}P_0$}. Yet, at the pore-scale the dynamics is still controlled by diffusion and the closure variables   $\boldsymbol\chi$ \deleted{and $\lambda$ are the solutions} \added{is the solution} of the closure problem\deleted{~\eqref{eq:closure-lambda-diffusive} and}~\eqref{eq:closure-chi-diffusive}.

\paragraph{Diffusion and advection dominate reaction} $\mbox{Da}<\omega$.
In this regime, reaction can be neglected compared to diffusive processes at the macroscale and the upscaled equation simplifies to 
\begin{linenomath*}
	\begin{align}
		\phi\dfrac{\partial \langle c\rangle_{\mathcal{IB}}}{\partial t}=\nabla\cdot\left[\tilde{\tilde{\textbf{D}}}^\star\nabla\langle c\rangle_{I\mathcal B}-\mbox{Pe}\langle c\rangle_{\mathcal{IB}}\langle\mathbf v\rangle_{\mathcal{IB}}\right],\quad (\mathbf x,t)\in \Omega\times(0,T), \label{eq:slowly positive alpha nonreacting}			\end{align}
		\end{linenomath*}
where $\tilde{\tilde{\textbf{D}}}^\star=\langle\textbf{D}(\mathbf I+\omega^{1-\gamma}\nabla_{\mathbf y}\boldsymbol{\chi})\rangle+\omega^{1-\alpha}\langle\boldsymbol{\chi}\mathbf k\rangle\cdot\nabla_{\mathbf x}P_0$, \deleted{$\tilde{\tilde{\textbf{D}}}'=\omega^{1-\gamma}\langle\textbf{D}\nabla_{\mathbf y}\lambda\rangle+\omega^{1-\alpha}\langle\lambda\mathbf k\rangle\cdot\nabla_{\mathbf x}P_0$ and  $\lambda$} and $\boldsymbol\chi$ still satisfies \eqref{eq:closure-chi-diffusive}.
\subsection{Moderately Fluctuating Boundary Conditions: $\omega <\varepsilon<\omega^{1/2}$}\label{moderately}
For this case, $\alpha$ always lies in $0\leq\alpha<1$ range.  Advection at the macroscale is non-negligible and the transport equation at the macroscale is described by Eq.\eqref{eq:upscaled eq}. 
The closure problems for \deleted{$\lambda$ and} $\boldsymbol\chi$ reduces to %
\begin{linenomath*}
\begin{subequations}\label{eq:closure-chi-steady}
	\begin{align}
&	\omega^{-\alpha}(\mathbf v_0-\langle \mathbf v_0\rangle)-\omega^{-\gamma}\nabla_{\mathbf y}\cdot\textbf{D}(\mathbf I+\omega^{1-\gamma}\nabla_{\mathbf y}\boldsymbol{\chi})+\omega^{1-\gamma-\alpha}\mathbf v_0\cdot(\nabla_{\mathbf y}\boldsymbol{\chi})=0, \quad \mathbf y\in \mathcal B,\\
&		\mathbf n\cdot\textbf{D} (\mathbf I+\omega^{1-\gamma}\nabla_{\mathbf y}\boldsymbol{\chi})=0, \quad \mathbf y\in \mathcal B,
	\end{align}
	\end{subequations}
\end{linenomath*} 
since the unsteady term can be neglected compared to diffusion and advection. In this regime, the  characteristic time scale of boundary fluctuations is much larger than  both diffusive and advective time scales. This results in a steady-state closure problem.

\paragraph{Diffusion and Advection Dominate Reaction} $\mbox{Da}<\omega$.
In this regime reaction is negligible and the upscaled equation \eqref{eq:upscaled eq} simplifies to \eqref{eq:slowly positive alpha nonreacting}	
where $\tilde{\tilde{\textbf{D}}}^\star=\langle\textbf{D}(\mathbf I+\omega^{1-\gamma}\nabla_{\mathbf y}\boldsymbol{\chi})\rangle+\omega^{1-\alpha}\langle\boldsymbol{\chi}\mathbf k\rangle\cdot\nabla_{\mathbf x}P_0$, \deleted{$\tilde{\tilde{\textbf{D}}}'=\omega^{1-\gamma}\langle\textbf{D}\nabla_{\mathbf y}\lambda\rangle+\omega^{1-\alpha}\langle\lambda\mathbf k\rangle\cdot\nabla_{\mathbf x}P_0$ and  $\lambda$} and $\boldsymbol\chi$  satisfies \deleted{\eqref{eq:closure-lambda-steady} and} \eqref{eq:closure-chi-steady}. \deleted{, respectively.}

\subsection{Highly Fluctuating Boundary Conditions: $\omega^{1/2}<\varepsilon<1$}\label{highly}
\subsubsection{Transport regime with $\mbox{Pe}<1$}
In this regime, the advective term at  the macro-scales is negligible. As a result the upscaled equation simplifies to Eq.~\eqref{eq:slowly negative alpha},
\begin{linenomath*}
	\begin{align}
		\phi\dfrac{\partial \langle c\rangle_{\mathcal{IB}}}{\partial t}=\nabla\cdot\left[\tilde{\tilde{\textbf{D}}}^\star\nabla\langle c\rangle_{\mathcal {IB}}\right]+\phi \omega^{-\gamma}\mathcal{K}^{\star}\mbox{Da}(1-\langle c\rangle_{\mathcal{IB}}^a),\nonumber			\end{align}
\end{linenomath*}
with  $\tilde{\tilde{\textbf{D}}}^\star=\langle\textbf{D}(\mathbf I+\omega^{1-\gamma}\nabla_{\mathbf y}\boldsymbol{\chi})\rangle$. \deleted{and $\tilde{\tilde{\textbf{D}}}'=\langle\textbf{D}\omega^{1-\gamma}\nabla_{\mathbf y}\lambda\rangle$} The closure variable \deleted{$\lambda$ and} $\boldsymbol \chi$ satisfies \added{instead an unsteady closure problem where unsteady effects, diffusion and advection are equally important, i.e.}
\begin{linenomath*}
	\begin{align}
		&\dfrac{\partial\boldsymbol{\chi}}{\partial \tau}-\omega^{-\gamma}\nabla_{\mathbf y}\cdot\textbf{D}(\mathbf I+\omega^{1-\gamma}\nabla_{\mathbf y}\boldsymbol{\chi})+\omega^{-\alpha}(\mathbf v_0-\langle \mathbf v_0\rangle)=0,\qquad\mathbf y\in \mathcal B,\\
		&-\mathbf n\cdot\textbf{D} (\mathbf I+\omega^{1-\gamma}\nabla_{\mathbf y}\boldsymbol{\chi})=0,\qquad\mathbf y \in\Gamma.
	\end{align}
\end{linenomath*}
\deleted{respectively.}
\paragraph{Diffusion and Advection Dominate Reaction} $\mbox{Da}<\omega$. In this regime ($\beta>1$) the reaction term at the macroscopic scale  is negligible and the upscaled equation is described by \eqref{eq:slowly positive alpha nonreacting}
where $\tilde{\tilde{\textbf{D}}}^\star=\langle\textbf{D}(\mathbf I+\omega^{1-\gamma}\nabla_{\mathbf y}\boldsymbol{\chi})\rangle$\deleted{and $\tilde{\tilde{\textbf{D}}}'=\langle\textbf{D}\omega^{1-\gamma}\nabla_{\mathbf y}\lambda\rangle$}.

\subsubsection{Transport regime with $1\leq\mbox{Pe}<\omega^{-1}$}
At the macroscale, dispersive and advective fluxes are of the same order of magnitude and the upscaled transport equation is given by Eq.\eqref{eq:upscaled eq} with  effective coefficients  defined by  Eqs.~\eqref{eq:dispersion D star}. Yet, diffusion is now negligible in the closure problem for \deleted{$\lambda$ and} $\boldsymbol\chi$, i.e.
\begin{linenomath*}
	\begin{align}
		&\dfrac{\partial \chi}{\partial \tau}+\omega^{-\alpha}(\mathbf v_0-\langle\mathbf v_0\rangle)+\omega^{1-\gamma-\alpha}\mathbf v_0\cdot(\nabla_{\mathbf y}\boldsymbol{\chi})=0,\qquad \mathbf y\in\mathcal B,\notag\\
		&-\mathbf n\cdot\textbf{D} (\mathbf I+\omega^{1-\gamma}\nabla_{\mathbf y}\boldsymbol{\chi})=0,\qquad\mathbf y \in{\Gamma},
	\end{align}
\end{linenomath*}
%
\paragraph{Diffusion and Advection Dominate Reaction}  $\mbox{Da}<\omega$.
In this regime the  reaction term at the macroscale is negligible, and the upscaled equation is given by ~Eq.\eqref{eq:upscaled eq} where the effective parameter $\tilde{\textbf{D}}^\star$ \deleted{and $\tilde{\textbf{D}}'$ are} \added{is} defined as \deleted{follows:}
 $\tilde{\textbf{D}}^\star=\langle\textbf{D}(\mathbf I+\omega^{1-\gamma}\nabla_{\mathbf y}\boldsymbol{\chi})\rangle+\omega^{1-\alpha}\langle\boldsymbol{\chi}\mathbf k(\mathbf y)\rangle\cdot\nabla_{\mathbf x}P_0$ \deleted{and 
$\tilde{\textbf{D}}'=\langle\textbf{D}\omega^{1-\gamma}\nabla_{\mathbf y}\lambda\rangle+\omega^{1-\alpha}\langle\lambda\mathbf k(\mathbf y)\rangle\cdot\nabla_{\mathbf x}P_0$.}
%

\section{Conclusion}\label{sec:conclusions}
Given the temporal variability of boundary conditions and forcings driving many subsurface processes, e.g. precipitation-driven transport in the vadose zone of arid and semiarid regions, or microbial activity and carbon cycling in the subsurface interaction zone (SIZ) controlled by seasonal mixing of surface water and groundwater in riverine systems, we investigate the impact of space-time averaging on nonlinear reactive transport in porous media. We are specifically concerned with understanding the impact of space-time upscaling in nonlinear systems driven by time-varying boundary conditions whose frequency is much larger than the characteristic time scale at which transport is studied or observed at the macroscopic scale.  Such systems are more vulnerable to upscaling approximations since the typical temporal resolution used in modern simulations significantly exceeds characteristic scales at which the system is driven.

We start by introducing the concept of spatiotemporal upscaling in the context of multiple-scale expansions. We then homogenize the pore-scale equations in space and time, and obtain a macroscopic equation which is dependent on the boundary condition frequency $\omega^{-1}$ and the  geometric separation of  scale parameter $\varepsilon$. Importantly, three different dynamical regimes are identified depending on the ratio between the diffusive time at the pore-scale ($\sim \varepsilon^2$) and the characteristic dimensionless period of the boundary temporal oscillations ($\omega$). They are referred to as slowly, moderately and highly fluctuating regimes. In the slowly fluctuating regime (when $\varepsilon\ll \omega$) pore-scale mass transport is entirely controlled by diffusion (and advection), and the local problem is steady state. In the highly fluctuating regime (when $\omega\ll \varepsilon$), pore-scale mass transport is affected by the additional time scale imposed by the boundary conditions and the local problem becomes unsteady. We refer to the moderately fluctuating regime if the period of the boundary conditions is comparable to the pore-scale diffusion time scale. \added{This analysis (i) supports the proposed classification in  three dynamical regimes, where the `speed of the fluctuation' (slow, moderate or high) is  quantified relatively to the characteristic diffusion time at the pore-scale, and (ii) provides insights on the primary mechanisms controlling mixing at the pore-scale.} We also identify the conditions under which scales are separable for any arbitrary $\omega$. Such conditions are  expressed in terms of the Pecl\'{e}t, Damk\"{o}hler numbers and the product between the boundary frequency $\omega^{-1} $and $\varepsilon$.

To conclude, the effects of lack of temporal resolution (i.e. temporal averaging) on nonlinear reactive transport  driven by time-varying boundary conditions or forcings should be accounted for at the macroscopic scale. The upscaling errors introduced by temporal (and spatial) averaging could have important implications especially when simulating systems for long temporal scales, i.e. when the observation time is much larger than the characteristic period of the oscillations.

%




\appendix
\section{Homogenization of the Transport Equation}\label{AppendixA}
As discussed in \cite{rajabi2021stochastic}, we present derivation of the upscaling procedure using space-time homogenization scheme. We start the upscaling procedure with the dimensionless pore-scale equation describing the transport of  the scalar function  $c_\omega(\textbf x,t)$  in an incompressible steady-state velocity field $\textbf{v}(\textbf{x})$,
\begin{linenomath*}
	\begin{align}\label{eq:transport_porescale}
		 \dfrac{\partial c_\omega}{\partial t}+\nabla \cdot(-\textbf D\nabla c_\omega+\mbox{Pe} \textbf{v} c_\omega)=0,\quad(\textbf x,t)\in{\Omega}^\omega_p\times(0,T),
		\end{align}
\end{linenomath*}
subject to the following boundary and initial conditions
\begin{linenomath*}	
\begin{align}	
		-\textbf n \cdot \textbf{D}\nabla c_\omega&=\mbox{Da}(c^a_\omega-1),& \textbf x\in\Gamma^\omega,\quad t>0,\label{eq:BC_porescale}\\
		 c_\omega(\textbf x,t=0)&=c_{\tiny\mbox{in}}(\textbf x), & \textbf  x\in \Omega_p^\omega.\label{IC}
	\end{align}
\end{linenomath*}
We define 
\begin{linenomath*}
	\begin{align} \label{eq:def}
	& t=\omega\tau,\qquad \textbf x=\varepsilon \textbf y,\qquad \varepsilon=\omega^\gamma,\qquad \mbox{Pe}=\omega^{-\alpha}, \qquad \mbox{Da}=\omega^\beta, 
	\end{align}
\end{linenomath*}
where $\mathbf y$ and  $\tau$  are the fast variables in space and time, respectively, and $\varepsilon\ll 1$ and $\omega\ll1$ are the spatial and temporal scale separation parameters. The exponents $\alpha$, $\beta$ and $\gamma$ identify the system's physical regimes. Particularly, $\gamma$ allows to represent the relationship between the frequency of boundary-imposed temporal fluctuations and the spatial heterogeneity. It is worth noticing that $\gamma>0$ since $\varepsilon \ll1$ and $\omega\ll1$. We first represent $c_\omega(\mathbf x,t)$ as $c_\omega(\mathbf x,t):=c(\textbf x,\textbf y,t,\tau)$. Given \eqref{eq:def}, the following relations hold for any space and time derivative in \eqref{eq:transport_porescale} \citep{rajabi2021stochastic},
\begin{linenomath*}
\begin{subequations}\label{eq:derivative}
	\begin{align}
	\dfrac{\partial c_\omega}{\partial t}&=\dfrac{\partial c}{\partial t}+\omega^{-1}\dfrac{\partial c}{\partial \tau},\\
	\nabla c_\omega&=\nabla_\mathbf x c+\varepsilon^{-1}\nabla_\mathbf y c.
	\end{align}
	\end{subequations}
\end{linenomath*}
Inserting  \eqref{eq:derivative}  into \eqref{eq:transport_porescale} leads to 
\begin{linenomath*}
	\begin{align}\label{intermediate1}
	\left(\dfrac{\partial c_\omega}{\partial t}+\omega^{-1}\dfrac{\partial c_\omega}{\partial \tau}\right)&+\nabla_\mathbf x\cdot[-\textbf{D}(\nabla_\mathbf x c_\omega+\varepsilon^{-1}\nabla_\mathbf yc_\omega)+\mbox{Pe} \mathbf v c_\omega]\nonumber \\&+\varepsilon^{-1}\nabla_\mathbf y\cdot[-\textbf{D}(\nabla_\mathbf xc_\omega+\varepsilon^{-1}\nabla_\mathbf y c_\omega)+\mbox{Pe}\mathbf v c_\omega]=0.	\end{align}
\end{linenomath*}
Expanding \eqref{intermediate1} up to order $\mathcal O(\omega^2)$, while using the \emph{ansatz} \eqref{asymp_exp}  \added{and the definitions \eqref{eq:def} for $\varepsilon$ and $\mbox{Pe}$, one obtains}
%
\deleted{Using definitions \eqref{eq:def} for $\varepsilon$ and $\mbox{Pe}$, one obtains}
\begin{linenomath*}
	\begin{align}
	\left(\dfrac{\partial c_0}{\partial t}\right.&\left.+\dfrac{1}{\omega}\dfrac{\partial c_0}{\partial \tau}\right)+\left(\omega\dfrac{\partial c_1}{\partial t}+\dfrac{\partial c_1}{\partial \tau}\right)+\left(\omega^2\dfrac{\partial c_2}{\partial t}+\omega\dfrac{\partial c_2}{\partial \tau}\right)\nonumber\\
	&-\nabla_{\mathbf x}\cdot\textbf{D}[\nabla_{\mathbf x} c_0 +\omega^{-\gamma}\nabla_{\mathbf y}c_0+\omega\nabla_{\mathbf x}c_1+\omega^{1-\gamma}\nabla_{\mathbf y}c_1+\omega^2\nabla_{\mathbf x}c_2+\omega^{2-\gamma}\nabla_{\mathbf y}c_2]\nonumber\\
	&+\nabla_{\mathbf x}\cdot[\omega^{-\alpha}\mathbf v_0c_0+\omega^{1-\alpha} (\mathbf v_0c_1+\mathbf v_1c_0)+\omega^{2-\alpha}(\mathbf v_0c_2+c_1\mathbf v_1+\mathbf v_2c_0)]\nonumber\\
	&-\nabla_{\mathbf y}\cdot\textbf{D}[\omega^{-\gamma}\nabla_{\mathbf x} c_0 +\omega^{-2\gamma}\nabla_{\mathbf y}c_0+\omega^{1-\gamma}\nabla_{\mathbf x}c_1+\omega^{1-2\gamma}\nabla_{\mathbf y}c_1+\omega^{2-\gamma}\nabla_{\mathbf x}c_2+\omega^{2-2\gamma}\nabla_{\mathbf y}c_2]\nonumber\\
	&+\nabla_{\mathbf y}\cdot[\omega^{-\alpha-\gamma}\mathbf v_0c_0+\omega^{1-\alpha-\gamma} (\mathbf v_0c_1+\mathbf v_1c_0)+\omega^{2-\alpha-\gamma}(\mathbf v_0c_2+c_1\mathbf v_1+\mathbf v_2c_0)]=0.
	\end{align}
\end{linenomath*}
%
We collect terms of like-powers of $ \omega$ as follows 
\begin{linenomath*}
\begin{align}\label{eq:sixteen}
\omega^{-1}&\left\{\dfrac{\partial c_0}{\partial \tau}-\omega^{1-2\gamma}\nabla_{\mathbf y}\cdot(\textbf{D}\nabla_{\mathbf y}c_0)+\omega^{1-\gamma-\alpha}\nabla_{\mathbf y}\cdot(c_0\mathbf v_0)\right\}+\nonumber\\
\omega^0&\left\{\left(\dfrac{\partial c_0}{\partial t}\right.+\dfrac{\partial c_1}{\partial \tau}\right)-\nabla_{\mathbf x}\cdot(\textbf{D}\nabla_{\mathbf x}c_0)-\omega^{-\gamma}[\nabla_{\mathbf x}\cdot(\textbf{D}\nabla_{\mathbf y}c_0)+\nabla_{\mathbf y} \cdot ( \textbf{D}\nabla_{\mathbf x}c_0)]-\omega^{1-2\gamma}\nabla_{\mathbf y}\cdot(\textbf{D}\nabla_{\mathbf y}c_1)+\nonumber\\
& \left.+ \omega^{-\alpha}\nabla_{\mathbf x}\cdot(c_0\mathbf v_0)+\omega^{1-\gamma-\alpha}\nabla_{\mathbf y}\cdot(\mathbf v_0c_1+\mathbf v_1c_0)\right\}+\nonumber\\
\omega&\left\{\left(\dfrac{\partial c_1}{\partial t}\right.+\dfrac{\partial c_2}{\partial \tau}\right)-\nabla_{\mathbf x}\cdot\textbf{D}(\nabla_{\mathbf x}c_1)-\omega^{-\gamma}[\nabla_{\mathbf x}\cdot(\textbf{D}\nabla_{\mathbf y}c_1)+\nabla_{\mathbf y}\cdot(\textbf{D}\nabla_{\mathbf x}c_1)]-\omega^{1-2\gamma}\nabla_{\mathbf y}\cdot(\textbf{D}\nabla_{\mathbf y}c_2)+\nonumber\\
&\left.+\omega^{-\alpha}\nabla_{\mathbf x}\cdot (\mathbf v_0c_1+\mathbf v_1c_0)+\omega^{1-\gamma-\alpha}\nabla_{\mathbf y}\cdot(\mathbf v_0c_2+\mathbf v_1c_1+\mathbf v_2c_0)\right\}=\mathcal{O}(\omega^2).
\end{align}
\end{linenomath*}
%
Similarly,  boundary condition \eqref{eq:BC_porescale} can be written as
\begin{linenomath*}
\begin{align}
-\mathbf n\cdot\textbf{D}(\nabla_{\mathbf x} c_0 +\varepsilon^{-1}\nabla_{\mathbf y}c_0+\omega\nabla_{\mathbf x}c_1+\omega\varepsilon^{-1}\nabla_{\mathbf y}c_1+\omega^2\nabla_{\mathbf x}c_2+\omega^2\varepsilon^{-1}\nabla_{\mathbf y}c_2)=\omega^{\beta}(c_0^a+a\omega c_0^{a-1}c_1-1).
\end{align}
\end{linenomath*}
Collecting terms of like-powers of $\omega$ one obtains
\begin{linenomath*}
\begin{align}
\omega^{-1}&[-\mathbf n\cdot(\omega^{1-\gamma}\textbf{D}\nabla_{\mathbf y}c_0)]+\omega^0[-\mathbf n\cdot \mathbf D(\nabla_{\mathbf x}c_0+\omega^{1-\gamma}\nabla_{\mathbf y}c_1)-\omega^{\beta}(c_0^a-1)]+\nonumber\\
\omega&[-\mathbf n\cdot \textbf{D}(\nabla_{\mathbf x}c_1+\omega^{1-\gamma}\nabla_{\mathbf y}c_2)-\omega^{\beta}ac_0^{a-1}c_1]=\mathcal{O}(\omega^2).\label{eq:BC-cascade}
\end{align}
\end{linenomath*}

\subsection{Terms of Order $\mathcal{O}(\omega^{-1})$}
At the leading order, \eqref{eq:sixteen} and \eqref{eq:BC-cascade} provide the following equation for $c_0$
\begin{linenomath*}
\begin{align}\label{eq:a12}
\dfrac{\partial c_0}{\partial \tau}-\omega^{1-2\gamma}\nabla_{\mathbf y}\cdot(\textbf{D}\nabla_{\mathbf y}c_0)+\omega^{1-\gamma-\alpha}\nabla_{\mathbf y}\cdot(c_0\mathbf v_0)=0,\quad \mathbf y \in\Omega_p, \, \tau\in \textbf{I}
\end{align}
\end{linenomath*}
subject to
\begin{linenomath*}
\begin{align}\label{eq:a13}
-\mathbf n\cdot(\omega^{1-\gamma}\textbf{D}\nabla_{\mathbf y}c_0)=0,\quad \mathbf y\in \Gamma,
\end{align}
\end{linenomath*}
i.e. $c_0=c_0(\mathbf x, t,\tau)$ since \eqref{eq:a12} and \eqref{eq:a13} are homogeneous. Integrating  \eqref{eq:a12} over $\Omega_p$ while applying the divergence theorem, one can write
\begin{linenomath*}
	\begin{align}
	\int_{\Omega_p}^{}\dfrac{\partial c_0}{\partial \tau} \mathrm d \mathbf y -\omega^{1-2\gamma}\int_{\Gamma}\mathbf{n}\cdot(\textbf{D}\nabla_{\mathbf y}c_0) \mathrm d\mathbf y+\omega^{1-\gamma-\alpha}\int_{\Gamma}\mathbf n\cdot(c_0\mathbf v_0) \mathrm d\mathbf y=0.\nonumber
	\end{align}
\end{linenomath*}
Accounting for \eqref{eq:a13} and the no-slip condition yields to 
\begin{linenomath*}
	\begin{align}
\int_{\Omega_p}\dfrac{\partial c_0}{\partial \tau} \mathrm d\mathbf y=0.\nonumber
	\end{align}
\end{linenomath*}
Since $\dfrac{\partial c_0}{\partial \tau}\geq 0 $, then $\dfrac{\partial c_0}{\partial \tau}=0$, i.e.  $c_0=c_0(\mathbf x, t)$.
\subsection{Terms of Order $\mathcal{O}(\omega^0)$}
Rearranging  \eqref{eq:sixteen} and \eqref{eq:BC-cascade} give
\begin{linenomath*}
\begin{align}\label{eq:order0}
&\left(\dfrac{\partial c_0}{\partial t}+\dfrac{\partial c_1}{\partial \tau}\right)-\nabla_{\mathbf x}\cdot(\textbf{D}\nabla_{\mathbf x}c_0)-\omega^{-\gamma}[\nabla_{\mathbf x}\cdot(\textbf{D}\nabla_{\mathbf y}c_0)]-\omega^{-\gamma}\nabla_{\mathbf y} \cdot [ \textbf{D}(\nabla_{\mathbf x}c_0+\omega^{1-\gamma}\nabla_{\mathbf y}c_1)]+\nonumber\\
& + \omega^{-\alpha}\nabla_{\mathbf x}\cdot(c_0\mathbf v_0)+\omega^{1-\gamma-\alpha}\nabla_{\mathbf y}\cdot(\mathbf v_0c_1+\mathbf v_1c_0)=0,
\end{align}
\end{linenomath*}
subject to
\begin{linenomath*}
\begin{align}\label{eq:BCorder0}
-\mathbf n\cdot \mathbf D(\nabla_{\mathbf x}c_0+\omega^{1-\gamma}\nabla_{\mathbf y}c_1)-\omega^{\beta}(c_0^a-1)=0,\quad \mathbf y\in \Gamma.
\end{align}
\end{linenomath*}
Integrating  \eqref{eq:order0} with respect to $\mathbf y$ and $\tau$ over $\mathcal B$ and $\mathcal I$, respectively, while noting that $\nabla_{\mathbf y}c_0\equiv 0$, and accounting for the divergence theorem and the boundary condition \eqref{eq:BCorder0}, leads to
\begin{linenomath*}
\begin{align}\label{order0integrated}
\dfrac{\partial c_0}{\partial t}=-\left\langle\dfrac{\partial c_1}{\partial \tau}\right\rangle_{\mathcal{IB}}+\nabla_{\mathbf x}\cdot(\textbf{D}\nabla_{\mathbf x}\langle c_0\rangle_{\mathcal{IB}})-\omega^{-\alpha}\nabla_{\mathbf x}\cdot(c_0\langle \mathbf v_0\rangle_{\mathcal{IB}})-\mathcal{K}^{\star}\omega^{\beta-\gamma}(c_0^a-1),
\end{align}
\end{linenomath*}
where $\mathcal{K}^{\star}=|\Gamma|/|\mathcal B|$.  Inserting  \eqref{order0integrated} into  \eqref{eq:order0} leads to
\begin{linenomath*}
\begin{align}\label{eq:17}
&\dfrac{\partial c_1}{\partial \tau}-\left\langle\dfrac{\partial c_1}{\partial \tau}\right\rangle_{\mathcal{IB}}-\omega^{-\alpha}\nabla_{\mathbf x}\cdot(c_0\langle \mathbf v_0\rangle_{\mathcal{IB}})-\mathcal{K}^{\star}\omega^{\beta-\gamma}(c_0^a-1)+ \omega^{-\alpha}\nabla_{\mathbf x}\cdot(c_0\mathbf v_0)\nonumber\\
&-\omega^{-\gamma}\nabla_{\mathbf y} \cdot [ \textbf{D}(\nabla_{\mathbf x}c_0+\omega^{1-\gamma}\nabla_{\mathbf y}c_1)]  +\omega^{1-\gamma-\alpha}\nabla_{\mathbf y}\cdot(\mathbf v_0c_1+\mathbf v_1c_0)=0,
\end{align}
\end{linenomath*}
since $c_0=\langle c_0\rangle_{\mathcal{IB}}$. Equation \eqref{eq:17}   is subject to \eqref{eq:BCorder0}. We look for a solution for $c_1(\mathbf x,t,\mathbf y,\tau)$ in the following form
\begin{linenomath*}
\begin{align}\label{closure}
c_1(\mathbf x,t,\mathbf y,\tau)=\boldsymbol \chi(\mathbf y,\tau)\cdot \nabla_{\mathbf x}c_0+ \lambda(\mathbf y,\tau)\dfrac{\partial c_0}{\partial t}+\overline c_1(\mathbf x,t),
\end{align}
\end{linenomath*}
where $\boldsymbol \chi(\mathbf y,\tau)$ and $\lambda(\mathbf y,\tau)$ are two unknown vector and scalar functions, and $\overline c_1(\mathbf x,t)$ is an integration function, respectively. \added{We emphasize that for `early' and `pre-asymptotic' times, i.e. when neither  time- or length-scales can be separated, or when no time-constraints are applicable but there is a separation of characteristic length scales, respectively, the postulated closure \eqref{closure} should, at least, exhibit memory effects (see e.g. \citep{parada-2011-Frequency,parada-2011-reply,wood2013volume}). Here, however, we are interested in long times, \emph{aka} `quasi-steady' state where both time- and spatial scales can be separated and local (in space and time) equations can be formulated.} Inserting \eqref{closure} into \eqref{eq:17} and \eqref{eq:BCorder0}, while noticing that \added{$\nabla_{\mathbf y}\cdot \mathbf v_0\equiv0$ and $\nabla_{\mathbf x}\cdot \langle\mathbf v_0\rangle\equiv0$ \citep{auriault1995taylor} and} $\partial_\tau \overline c_1=\nabla_{\mathbf y}\overline c_1\equiv 0$, gives
\begin{linenomath*}
\begin{align}\label{eq:20}
&\left[\dfrac{\partial \lambda}{\partial \tau}-\left\langle\dfrac{\partial \lambda}{\partial \tau}\right\rangle_{\mathcal{IB}}-\omega^{1-2\gamma}\nabla_{\mathbf y}\cdot(\mathbf D \nabla_{\mathbf y}\lambda)+\omega^{1-\gamma-\alpha}\mathbf v_0\cdot\nabla_{\mathbf y}\lambda\right]\dfrac{\partial c_0}{\partial t}+\nonumber\\
&\left[\dfrac{\partial\boldsymbol \chi}{\partial \tau}-\left\langle\dfrac{\partial \boldsymbol\chi}{\partial \tau}\right\rangle_{\mathcal{IB}}-\omega^{-\alpha}\langle \mathbf v_0\rangle_{\mathcal{IB}}+\omega^{-\alpha}\mathbf v_0-\omega^{-\gamma}\nabla_{\mathbf y}\cdot[\mathbf D(\mathbf I+\omega^{1-\gamma}\nabla_{\mathbf y}\boldsymbol \chi)]+\omega^{1-\gamma-\alpha}\mathbf v_0\cdot\nabla_{\mathbf y}\boldsymbol\chi\right]\cdot\nabla_{\mathbf x}c_0+\nonumber\\
&\omega^{-\alpha}(\nabla_{\mathbf x} \cdot \mathbf v_0+\omega^{1-\gamma}\nabla_{\mathbf y}\cdot \mathbf v_1)c_0+\omega^{1-\gamma-\alpha}\mathbf v_0\cdot \nabla_{\mathbf y}\overline c_1-\mathcal{K}^\star\omega^{\beta-\gamma}(c_0^a-1)=0,
\end{align}
\end{linenomath*}
where $\mathbf I$ is the identity matrix. \deleted{Collecting terms, one obtains}
%
\deleted{since  $\nabla_{\mathbf y}\cdot \mathbf v_0\equiv0$ and $\nabla_{\mathbf x}\cdot \langle\mathbf v_0\rangle\equiv0$.} 
Equation \eqref{eq:20} is subject to the boundary condition
\begin{linenomath*}
\begin{align}\label{eq:21}
&-\mathbf n\cdot\textbf{D}\left[(\mathbf I+\omega^{1-\gamma}\nabla_{\mathbf y}\boldsymbol \chi)\cdot\nabla_{\mathbf x}c_0+\omega^{1-\gamma}\nabla_{\mathbf y}\lambda
\dfrac{\partial c_0}{\partial t}\right]=\omega^{\beta}(c_0^a-1).
\end{align}
\end{linenomath*}
Expanding the continuity equation $\nabla\cdot\mathbf{v}_\omega=\nabla_{\mathbf x}\cdot(\mathbf v_0+\omega\mathbf v_1+\omega^2\mathbf v_2)+\varepsilon^{-1}\nabla_{\mathbf y}\cdot(\mathbf v_0+\omega\mathbf v_1+\omega^2\mathbf v_2)=0$ leads to
\begin{linenomath*}
\begin{align}
\omega^{-1}(\omega^{1-\gamma}\nabla_{\mathbf y}\cdot \mathbf v_0)+\omega^0(\nabla_{\mathbf x}\cdot \mathbf v_0+\omega^{1-\gamma}\nabla_{\mathbf y}\cdot \mathbf v_1)+\omega(\nabla_{\mathbf x}\cdot \mathbf v_1+\omega^{1-\gamma}\nabla_{\mathbf y}\cdot \mathbf v_2)=\mathcal{O}(\omega^2)
\end{align}
\end{linenomath*}
i.e. $\nabla_{\mathbf x}\cdot \mathbf v_0+\omega^{1-\gamma}\nabla_{\mathbf y}\cdot \mathbf v_1==0$ and \eqref{eq:20} reduces to
\begin{linenomath*}
\begin{align}\label{ciao}
&\left[\dfrac{\partial\boldsymbol \chi}{\partial \tau}-\omega^{-\alpha}\langle \mathbf v_0\rangle_{\mathcal{IB}}+\omega^{-\alpha}\mathbf v_0-\omega^{-\gamma}\nabla_{\mathbf y}\cdot[\mathbf D(\mathbf I+\omega^{1-\gamma}\nabla_{\mathbf y}\boldsymbol \chi)]+\omega^{1-\gamma-\alpha}\mathbf v_0\cdot\nabla_{\mathbf y}\boldsymbol\chi\right]\cdot\nabla_{\mathbf x}c_0+\nonumber\\
&\left[\dfrac{\partial \lambda}{\partial \tau}-\omega^{1-2\gamma}\nabla_{\mathbf y}\cdot(\mathbf D \nabla_{\mathbf y}\lambda)+\omega^{1-\gamma-\alpha}\mathbf v_0\cdot\nabla_{\mathbf y}\lambda\right]\dfrac{\partial c_0}{\partial t}=\mathcal{K}^\star\omega^{\beta-\gamma}(c_0^a-1),
\end{align}
\end{linenomath*}
since $\langle \lambda\rangle_\mathcal{B}=\langle \boldsymbol \chi\rangle_\mathcal{B}=0$.
In order to decouple the pore-scale from the continuum-scale, \added{it is sufficient that the closure problem \eqref{ciao} is independent of macroscopic quantities, such as $\dfrac{\partial c_0}{\partial t}$ and $\nabla_{\mathbf x} c_0$. Therefore, one needs to impose that these terms are negligible relative to all others for all possible values of $\alpha$, $\beta$ and $\gamma$. This results on constraining the exponents in the coefficients multiplying these coupling terms. Specifically, in order to separate scales,} it is sufficient that
\begin{linenomath*}
	\begin{align}\label{decoupling_condition}
	&\beta-\gamma>M
		\end{align}
\end{linenomath*}
where
\begin{linenomath*}\label{eq:M}
	\begin{align}
M:=\max\lbrace0,-\gamma,1-\alpha-\gamma,-\alpha,1-2\gamma\rbrace.
\end{align}
\end{linenomath*}
 Additionally,  $\beta>\max\{0,1-\gamma\}$, i.e.
\begin{linenomath*}
	\begin{align}\label{decoupling_condition2}
	&\beta>0,
	\end{align}
\end{linenomath*}
since $\gamma>0$.  We emphasize that condition \eqref{decoupling_condition2} is automatically satisfied if  \eqref{decoupling_condition} is satisfied since both $\gamma>0$ and $M>0$. 
\added{Once the conditions under which scales are decoupled have been identified, appropriate initial conditions need to be formulated. We start by expanding Eq.~\eqref{IC} at $t=\tau=0$, i.e.  $c_\omega(\textbf x,t=0)=c_{\tiny\mbox{in}}(\textbf x)$}
\begin{linenomath*}
\begin{align}
c_{\tiny\mbox{in}}(\textbf x)=c_{0, \tiny\mbox{in}}(\textbf x)+\omega c_{1, \tiny\mbox{in}}(\textbf x, \mathbf y)=c_{0, \tiny\mbox{in}}(\textbf x)+\omega \left[ \boldsymbol \chi_{\tiny\mbox{in}}(\mathbf y)\cdot\left. \nabla_{\mathbf x}c_0\right|_{t=0}+ \lambda_{\tiny\mbox{in}}(\mathbf y)\left.\dfrac{\partial c_0}{\partial t}\right|_{t=0}+\overline c_1(\mathbf x, t=0) \right]
\end{align}
\end{linenomath*}
\added{At the leading order,  $c_{\tiny\mbox{in}}(\textbf x)=c_{0, \tiny\mbox{in}}$. At the order $\omega$,}
\begin{linenomath*}
\begin{align}\label{IC_compatibility}
\boldsymbol \chi_{\tiny\mbox{in}}(\mathbf y)\cdot\left. \nabla_{\mathbf x}c_0\right|_{t=0}+ \lambda_{\tiny\mbox{in}}(\mathbf y)\left.\dfrac{\partial c_0}{\partial t}\right|_{t=0}=0,
\end{align}
\end{linenomath*}
\added{if we set $c_1(\mathbf x, t=0)=0$. Since $\left.\nabla_{\mathbf x}c_0\right|_{t=0}$ and $\left.\dfrac{\partial c_0}{\partial t}\right |_{t=0}$ are known functions of $\mathbf x$, the compatibility condition \eqref{IC_compatibility} requires $\chi_{\tiny\mbox{in}}(\mathbf y)=\lambda_{\tiny\mbox{in}}(\mathbf y)=0$.} The former conditions allow one to write the following closure problems for $\boldsymbol \chi$ and $\lambda$, 
\begin{linenomath*}
\begin{align}
\dfrac{\partial\boldsymbol \chi}{\partial \tau}-\omega^{-\alpha}\langle \mathbf v_0\rangle_{\mathcal{IB}}+\omega^{-\alpha}\mathbf v_0-\omega^{-\gamma}\nabla_{\mathbf y}\cdot[\mathbf D(\mathbf I+\omega^{1-\gamma}\nabla_{\mathbf y}\boldsymbol \chi)]+\omega^{1-\gamma-\alpha}\mathbf v_0\cdot\nabla_{\mathbf y}\boldsymbol\chi=0,
\end{align}
\end{linenomath*}
subject to
\begin{linenomath*}
\begin{align}
&-\mathbf n\cdot \textbf{D} (\mathbf I+\omega^{1-\gamma}\nabla_{\mathbf y}\boldsymbol\chi)=0,\\
&\boldsymbol\chi(\mathbf y, \tau=0)=\boldsymbol\chi_{\tiny\mbox{in}}(\mathbf y)=0,
\end{align}
\end{linenomath*}
and
\begin{linenomath*}
\begin{align}
\dfrac{\partial \lambda}{\partial \tau}-\omega^{1-2\gamma}\nabla_{\mathbf y}\cdot(\mathbf D \nabla_{\mathbf y}\lambda)+\omega^{1-\gamma-\alpha}\mathbf v_0\cdot\nabla_{\mathbf y}\lambda=0,
\end{align}
\end{linenomath*}
subject to
\begin{linenomath*}
\begin{align}
&-\mathbf n\cdot\textbf{D} \nabla_{\mathbf y}\lambda=0,\\
&\lambda(\mathbf y, \tau=0)=\lambda_{\tiny\mbox{in}}(\mathbf y)=0.
\end{align}
\end{linenomath*}
\added{It is important to note that the closure problem for $\lambda$ is homogeneous, i.e. the postulation for  $c_1$ for long times,  reduces to the classical closure}
\begin{linenomath*}
\begin{align}\label{closure}
c_1(\mathbf x,t,\mathbf y,\tau)=\boldsymbol \chi(\mathbf y,\tau)\cdot \nabla_{\mathbf x}c_0+\overline c_1(\mathbf x,t).
\end{align}
\end{linenomath*}
\added{i.e. $\lambda\equiv 0$.}


\subsubsection{Conditions}
In this section, we investigate how \eqref{decoupling_condition} translates into constraints on  $\alpha$ and $\beta$ for different values of $\gamma$. \added{We do so by hypothesizing the value of the maximum $M$, defined by \eqref{eq:M}, among the four possible scenarios: $M=0$, $M=1-\alpha-\gamma$, $M=-\alpha$ and $M=1-2\gamma$. We emphasize that once the physical system under study is identified both in terms of physical domain (i.e. $\epsilon$), boundary conditions (i.e. $\gamma$) and dynamic regimes ($\alpha$ and $\beta$), the parameters $\epsilon$, $\gamma$, $\alpha$ and $\beta$ are fixed, $M$ is a uniquely defined scalar, and \eqref{decoupling_condition} must be satisfied if scales are decoupled. If \eqref{decoupling_condition} is not satisfied, then \eqref{eq:upscaled eq} may not represent spatio-temporally averaged pore-scale processes with the accuracy prescribed by the homogenization procedure. In the following, we rewrite the applicability condition \eqref{decoupling_condition} in terms of Da and Pe, so that its ramification on dynamical regimes is made explicit. }
\paragraph{When $M=0$} Conditions \eqref{decoupling_condition} are reformulated as

\begin{linenomath*}
\begin{align}
\begin{cases}
\alpha>0\\
\gamma>1/2\\
\alpha>1-\gamma
\end{cases}
\Rightarrow \beta>\gamma, 
\end{align}
\end{linenomath*}
i.e. Da$<\varepsilon$.

\paragraph{When $M=1-\alpha-\gamma$} Conditions \eqref{decoupling_condition} are reformulated as

\begin{linenomath*}
\begin{align}
\begin{cases}
\alpha<\gamma\\
\gamma<1\\
\alpha<1-\gamma
\end{cases}
\Rightarrow \beta>1-\alpha, 
\end{align}
\end{linenomath*}
i.e. $\mbox{Da}/\mbox{Pe}<\omega$.

\paragraph{When $M=-\alpha$} Conditions \eqref{decoupling_condition} are reformulated as

\begin{linenomath*}
\begin{align}
\begin{cases}
\alpha<0\\
\gamma>1
\end{cases}
\Rightarrow \beta>\gamma-\alpha,
\end{align}
\end{linenomath*}
i.e. $\mbox{Da}/\mbox{Pe}<\varepsilon$.

\paragraph{When $M=1-2\gamma$} Conditions \eqref{decoupling_condition} are reformulated as

\begin{linenomath*}
\begin{align}
\begin{cases}
\alpha>\gamma\\
\gamma<1/2
\end{cases}
\Rightarrow \beta>1-\gamma,
\end{align}
\end{linenomath*}
i.e. $\mbox{Da}<\omega/\varepsilon$.

We emphasize that the case $M=-\gamma$ requires $\gamma<0$. This violates the assumption that $\gamma>0$. As a result, this case is not self-consistent with the homogenization procedure and should be ignored. 

\begin{figure}[!ht]
  	\centering
  	\includegraphics[width=1\textwidth]{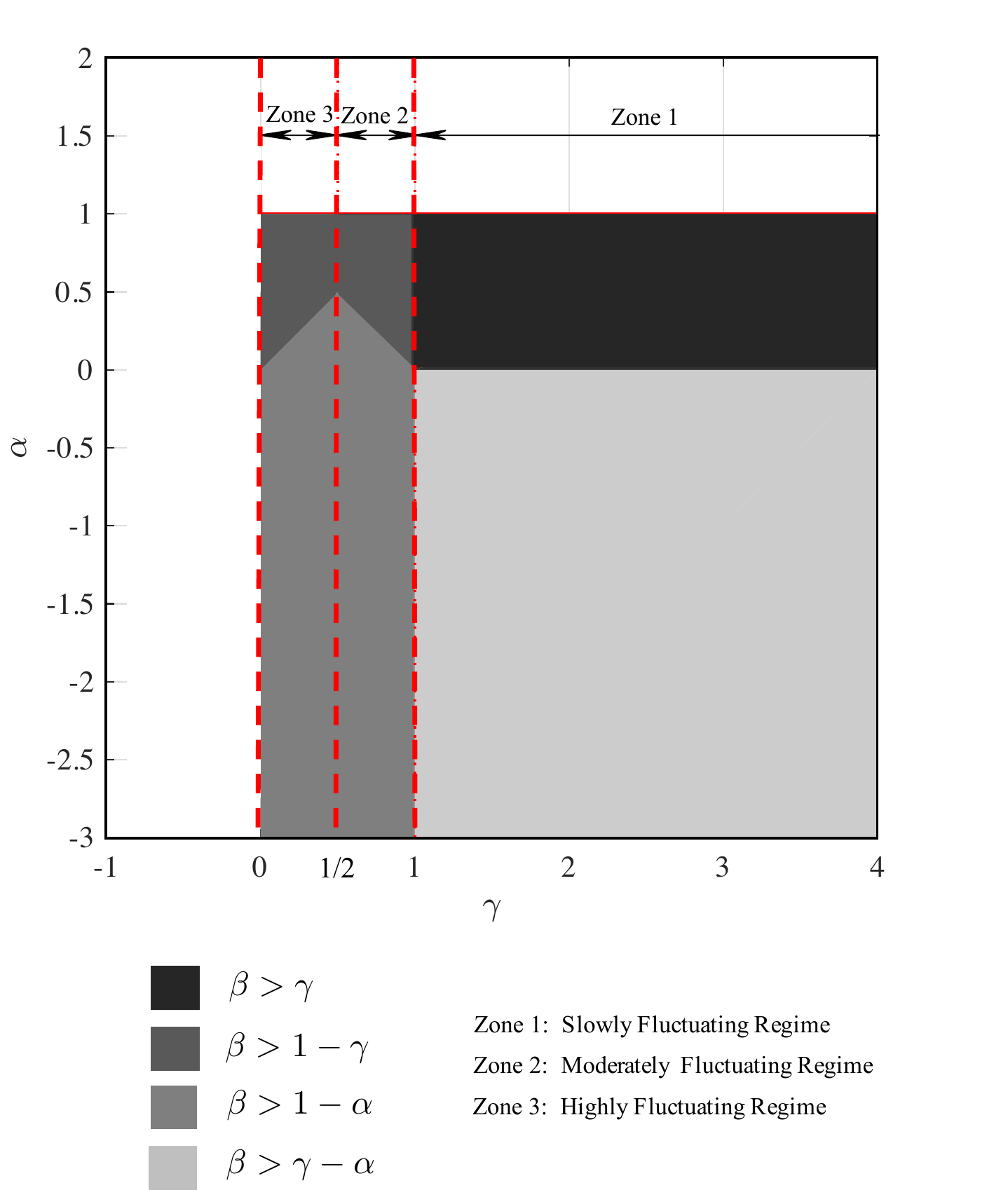}
  	\caption{Diagram in the ($\alpha,\gamma$)-space summarizing the relationship between the Pecl\'{e}t  $(=\omega^\alpha)$, Damk\"{o}hler $(=\omega^\beta)$ and the ratio between space and time scale parameters  $(\gamma=\mbox{log}\varepsilon/\mbox{log}\omega)$ in the three (slowly, moderately and highly fluctuating) regimes for a system with separation of scale parameter $\varepsilon\ll1$ and boundary condition frequency $1/\omega\gg 1$.}\label{allzones}
  \end{figure}

The previous conditions are summarized in the $(\alpha,\gamma)$-plane of Figure~\ref{allzones}. 


The system behavior can be classified based on the magnitude of $\gamma$:
\begin{itemize}
\item When $\gamma>1$, i.e. $\varepsilon<\omega$, the system is referred to as \emph{slowly fluctuating}; the conditions  to guarantee that scale separation occur are summarized in the $(\alpha,\beta)$-plane in the Figure~\ref{All_New}(a);
\item When $1/2<\gamma<1$, i.e. $\omega<\varepsilon<\omega^{1/2}$ (or $\omega\approx\varepsilon$), the system is referred to as \emph{moderately fluctuating}; the conditions  to guarantee that scale separation occur are summarized in the $(\alpha,\beta)$-plane in the Figure~\ref{All_New}(b);
\item When $0<\gamma<1/2$, i.e. $\omega^{1/2}<\varepsilon<1$ (or $\varepsilon\gg\omega$), the system is referred to as \emph{highly fluctuating}; the conditions  to guarantee that scale separation occur are summarized in the $(\alpha,\beta)$-plane in the Figure~\ref{All_New}(c).
\end{itemize}

\subsection{Terms of Order  $\mathcal{O}(\omega^1)$}
At the following order, we have
\deleted{Rearranging  \eqref{eq:a35} yields to}
\begin{linenomath*}
	\begin{align}\label{eq:a36}
&\left(\dfrac{\partial c_1}{\partial t}+\dfrac{\partial c_2}{\partial \tau}\right)-\nabla_{\mathbf x}\cdot(\textbf{D}\nabla_{\mathbf x}c_1)- \omega^{-\gamma}[\nabla_{\mathbf x}\cdot(\textbf{D}\nabla_{\mathbf y}c_1)]+\nonumber\\
&	-\omega^{-\gamma}\nabla_{\mathbf y}\cdot\textbf{D}(\nabla_{\mathbf x}c_1+\omega^{1-\gamma}\nabla_{\mathbf y}c_2)+ \omega^{-\alpha}\nabla_{\mathbf x}\cdot (\mathbf v_0c_1+\mathbf v_1c_0)+\omega^{1-\gamma-\alpha}\nabla_{\mathbf y}\cdot(\mathbf v_0c_2+\mathbf v_1c_1+\mathbf v_2c_0)=0
	\end{align}
\end{linenomath*}
subject to
\begin{linenomath*}
\begin{align}\label{bc:lastorder}
&-\mathbf n\cdot\mathbf{D}(\nabla_{\mathbf x}c_1+\omega^{1-\gamma}\nabla_{\mathbf y}c_2)-\omega^{\beta}ac_0^{a-1}c_1=0.
\end{align}
\end{linenomath*}

Integrating \eqref{eq:a36} over $\mathcal B$ and $\mathcal I$ with respect to $\mathbf y$ and $\tau$, while accounting for \eqref{closure}, $\langle \chi\rangle=0$, we obtain,
\begin{linenomath*}
	\begin{align}\label{eq:a38}
\left\langle\dfrac{\partial c_1}{\partial t}\right\rangle_{\mathcal{IB}}&+\left\langle\dfrac{\partial c_2}{\partial \tau}\right\rangle_{\mathcal{IB}}-\nabla_{\mathbf x}\cdot\left[\textbf{D}\nabla_{\mathbf x}\left(\left\langle\boldsymbol \chi(\mathbf y,\tau)\right\rangle_{\mathcal{IB}}\cdot \nabla_{\mathbf x}c_0+\overline c_1(\mathbf x,t)\right)\right]\nonumber \\
& - \omega^{-\gamma}\left[\nabla_{\mathbf x}\cdot\left\langle \textbf{D}\nabla_{\mathbf y}\left(\boldsymbol \chi(\mathbf y,\tau)\cdot \nabla_{\mathbf x}c_0+\overline c_1(\mathbf x,t)\right)\right\rangle_{\mathcal{IB}}\right]	\nonumber\\
&-\omega^{-\gamma}\left\langle\nabla_{\mathbf y}\cdot\textbf{D}(\nabla_{\mathbf x}c_1+\omega^{1-\gamma}\nabla_{\mathbf y}c_2)\right\rangle_{\mathcal{IB}}+\omega^{1-\gamma-\alpha}\left\langle\nabla_{\mathbf y}\cdot(\mathbf v_0c_2+\mathbf v_1c_1+\mathbf v_2c_0)\right\rangle_{\mathcal{IB}}\nonumber\\
&+\omega^{-\alpha}\nabla_{\mathbf x}\cdot \left\langle \mathbf v_0c_1+\mathbf v_1c_0\right\rangle_{\mathcal{IB}}=0
	\end{align}
\end{linenomath*}
The third term in \eqref{eq:a38} is identically equal to zero since $\langle \chi\rangle=0$ and the arbitrary integrating function $\overline c_1$ can be selected such that $\nabla_{\mathbf x}\cdot  (\mathbf D \nabla_{\mathbf x} \overline c_1)=0$, i.e. if $\overline c_1$ is linear in $\mathbf x$. Similarly, $\left\langle\nabla_{\mathbf y}\cdot(\mathbf v_0c_2+\mathbf v_1c_1+\mathbf v_2c_0)\right\rangle_{\mathcal{IB}}=0$ because of the divergence theorem, the no-slip boundary condition on $\Gamma$ and periodicity on the unit cell boundaries. Therefore,  \eqref{eq:a38} simplifies to
\begin{linenomath*}
	\begin{align}\label{eq:a39}
\left\langle\dfrac{\partial c_1}{\partial t}\right\rangle_{\mathcal{IB}}&+\left\langle\dfrac{\partial c_2}{\partial \tau}\right\rangle_{\mathcal{IB}} - \omega^{-\gamma}\left[\nabla_{\mathbf{x}}\cdot\left(\left\langle \textbf{D}\nabla_{\mathbf y}\boldsymbol \chi(\mathbf y,\tau)\right\rangle_{\mathcal{IB}}\cdot \nabla_{\mathbf x}c_0\right)\right]	\nonumber\\
&+\omega^{-\alpha}\nabla_{\mathbf x}\cdot \left\langle \mathbf v_0c_1+\mathbf v_1c_0\right\rangle_{\mathcal{IB}} -\omega^{-\gamma}\left\langle\nabla_{\mathbf y}\cdot\textbf{D}(\nabla_{\mathbf x}c_1+\omega^{1-\gamma}\nabla_{\mathbf y}c_2)\right\rangle_{\mathcal{IB}}=0.
	\end{align}
\end{linenomath*}
We proceed further by analyzing the last two terms separately. We start with the fourth term in \eqref{eq:a39}, $\nabla_{\mathbf x}\cdot \left\langle \mathbf v_0c_1+\mathbf v_1c_0\right\rangle_{\mathcal{IB}}$. Combining it with \eqref{closure} and $\mathbf v_0=-\mathbf k(\mathbf y)\cdot \nabla_{\mathbf x}P_0$ one obtains
\begin{linenomath*}
	\begin{align}\label{eq:order1_3rd_term_on_RHS}
	\nabla_{\mathbf x}\cdot \left\langle\mathbf v_0c_1+\mathbf v_1c_0\right\rangle_{\mathcal{IB}}=-\nabla_{\mathbf x}\cdot \left\langle\mathbf k\nabla_{\mathbf x}P_0\left(\boldsymbol\chi\cdot\nabla_{\mathbf x}c_0+\overline{c}_1\right)\right\rangle_{\mathcal{IB}}+\nabla_{\mathbf x}\cdot\left\langle\mathbf v_1c_0\right\rangle_{\mathcal{IB}}. 
	\end{align}
\end{linenomath*}
Using Einstein notation convention and indicial notation, one can write
\begin{linenomath*}
	\begin{align}\label{eq:41}
	\nabla_{\mathbf x}\cdot \left\langle\mathbf v_0c_1+\mathbf v_1c_0\right\rangle_{\mathcal{IB}}&=\frac{\partial}{\partial x_i}\left\langle v_{0i}c_1+ v_{1i}c_0\right\rangle_{\mathcal{IB}}\nonumber \\
	&=-\frac{\partial}{\partial x_i} \left\langle k_{ij}\frac{\partial P_0}{\partial x_j}\left(\chi_m\frac{\partial c_0}{\partial x_m}+\overline{c}_1\right)\right\rangle_{\mathcal{IB}}+\frac{\partial}{\partial x_i}\left\langle v_{1i}c_0\right\rangle_{\mathcal{IB}} \nonumber \\
	&=- \left\langle k_{ij}\chi_m\right\rangle_{\mathcal{IB}}\left(\frac{\partial^2 P_0}{\partial x_i\partial x_j}\frac{\partial c_0}{\partial x_m}+\frac{\partial P_0}{\partial x_j}\frac{\partial^2 c_0}{\partial x_i\partial x_m}\right)\nonumber\\
	&\quad\,-\left\langle  k_{ij}\right\rangle_{\mathcal{IB}}\frac{\partial}{\partial x_i}\left(\frac{\partial P_0}{\partial x_j}\overline{c}_1\right)+\frac{\partial}{\partial x_i}\left\langle v_{1i}c_0\right\rangle_{\mathcal{IB}}.
	\end{align}
\end{linenomath*}
Noticing that $\nabla_{\mathbf x}\cdot \langle\mathbf v_0\rangle_{\mathcal{IB}}\equiv0$, this results in
\begin{linenomath*}
	\begin{align}\label{eq:laplaceP}
	&\frac{\partial \langle v_{0i}\rangle_{\mathcal{IB}}}{\partial x_i} =-\frac{\partial }{\partial x_i} \left(\langle k_{ij}\rangle_{\mathcal{IB}}\frac{\partial P_0}{\partial x_j}\right)=-\langle k_{ij}\rangle_{\mathcal{IB}}\frac{\partial^2 P_0}{\partial x_i\partial x_j}\equiv 0\end{align}
\end{linenomath*}
i.e. $\partial_{x_ix_j}^2 P_0\equiv0$, since $\langle k_{ij}\rangle_{\mathcal{IB}}\neq 0$. Therefore, \eqref{eq:41} can be simplified as follows
\begin{linenomath*}
	\begin{align}\label{eq:43}
	\nabla_{\mathbf x}\cdot \left\langle\mathbf v_0c_1+\mathbf v_1c_0\right\rangle_{\mathcal{IB}}&=- \frac{\partial^2 c_0}{\partial x_i\partial x_m}\left\langle\chi_m k_{ij}\right\rangle_{\mathcal{IB}}\frac{\partial P_0}{\partial x_j}-\frac{\partial}{\partial x_i}\left(\left\langle  k_{ij}\right\rangle_{\mathcal{IB}}\frac{\partial P_0}{\partial x_j}\overline{c}_1\right)+\frac{\partial}{\partial x_i}\left\langle v_{1i}c_0\right\rangle_{\mathcal{IB}}\nonumber \\
	&=- \left[\left\langle\boldsymbol\chi \mathbf k\right\rangle_{\mathcal{IB}}\cdot\nabla_{\mathbf x} P_0\right]_{mi}\frac{\partial}{\partial x_i}\left(\frac{\partial c_0}{\partial x_m}\right)\nonumber\\
	&\quad-\frac{\partial}{\partial x_i}\left(\left [\left\langle  \mathbf k \right\rangle_{\mathcal{IB}}\cdot \nabla_{\mathbf x} P_0\right]_i\overline{c}_1\right)+\frac{\partial}{\partial x_i}\left\langle v_{1i}c_0\right\rangle_{\mathcal{IB}} \nonumber\\
	&=- \left[\left(\left\langle\boldsymbol\chi \mathbf k\right\rangle_{\mathcal{IB}}\cdot\nabla_{\mathbf x} P_0\right)\cdot \nabla_{\mathbf x}\right]\cdot \nabla_{\mathbf x}c_0 \nonumber\\
	&\quad -\nabla_{\mathbf x}\cdot\left(\left\langle  \mathbf k \right\rangle_{\mathcal{IB}}\cdot \nabla_{\mathbf x} P_0\overline{c}_1\right)+\nabla_{\mathbf x}\cdot\left(\left\langle \mathbf v_{1}\right\rangle_{\mathcal{IB}} c_0 \right)
	\end{align}
\end{linenomath*}
Using the divergence theorem and the boundary condition \eqref{bc:lastorder}, the last term in \eqref{eq:a39} can be written as
\begin{linenomath*}
\begin{align}\label{eq:order1_4rd_term_on_RHS}
&\omega^{-\gamma}\left\langle\nabla_{\mathbf y}\cdot\textbf{D}(\nabla_{\mathbf x}c_1+\omega^{1-\gamma}\nabla_{\mathbf y}c_2)\right\rangle_{\mathcal{IB}}=-\omega^{\beta-\gamma}\mathcal{K^\star}ac_0^{a-1}\langle c_1\rangle_{\mathcal I\Gamma},
\end{align}
\end{linenomath*}
where $\mathcal{K^\star}=\dfrac{|\Gamma|}{|\mathcal B|}$. Inserting \eqref{eq:43} and \eqref{eq:order1_4rd_term_on_RHS} in \eqref{eq:a39}, white noting that $\langle A\rangle=\phi\langle A\rangle_{\mathcal{IB}}$ and $\overline c_1=\langle c_1\rangle$, we obtain

\begin{linenomath*}
	\begin{align}\label{eq:45}
\left\langle\dfrac{\partial c_1}{\partial t}\right\rangle_{\mathcal{IB}}&+\left\langle\dfrac{\partial c_2}{\partial \tau}\right\rangle_{\mathcal{IB}} -\phi^{-1} \omega^{-\gamma}\left[\nabla_{\mathbf{x}}\cdot\left(\left\langle \textbf{D}\nabla_{\mathbf y}\boldsymbol \chi\right\rangle\cdot \nabla_{\mathbf x}c_0\right)\right]-\phi^{-1}\omega^{-\alpha}  \left[\left(\left\langle\boldsymbol\chi \mathbf k\right\rangle\cdot\nabla_{\mathbf x} P_0\right)\cdot \nabla_{\mathbf x}\right]\cdot \nabla_{\mathbf x}c_0\nonumber \\
&+\omega^{-\alpha} \nabla_{\mathbf x}\cdot\left(\left\langle  \mathbf v_0 \right\rangle_{\mathcal{IB}}\overline{c}_1\right)+\omega^{-\alpha} \nabla_{\mathbf x}\cdot\left(\left\langle \mathbf v_{1}\right\rangle_{\mathcal{IB}} c_0 \right) +\omega^{\beta-\gamma}\mathcal{K^\star}ac_0^{a-1}\langle c_1\rangle_{\mathcal I\Gamma}=0.
	\end{align}
\end{linenomath*}
Importantly, since $\left[\left(\left\langle\boldsymbol\chi \mathbf k\right\rangle\cdot\nabla_{\mathbf x} P_0\right)\cdot \nabla_{\mathbf x}\right]\cdot \nabla_{\mathbf x}c_0=\nabla_{\mathbf x}\cdot \left[\left(\left\langle\boldsymbol\chi \mathbf k\right\rangle\cdot\nabla_{\mathbf x} P_0\right)\cdot \nabla_{\mathbf x}c_0\right]$ because of \eqref{eq:laplaceP}, \eqref{eq:45} can be rearranged as follows
\begin{linenomath*}
	\begin{align}\label{eq:46}
\left\langle\dfrac{\partial c_1}{\partial t}\right\rangle_{\mathcal{IB}}&+\left\langle\dfrac{\partial c_2}{\partial \tau}\right\rangle_{\mathcal{IB}} -\phi^{-1} \omega^{-1}\nabla_{\mathbf{x}}\cdot\left[\left(\omega^{1-\gamma}\left\langle \textbf{D}\nabla_{\mathbf y}\boldsymbol \chi\right\rangle+ \omega^{1-\alpha}\left\langle\boldsymbol\chi \mathbf k\right\rangle\cdot\nabla_{\mathbf x} P_0\right)\cdot \nabla_{\mathbf x}c_0\right]\nonumber \\ 
&+\omega^{-\alpha} \nabla_{\mathbf x}\cdot\left(\left\langle  \mathbf v_0 \right\rangle_{\mathcal{IB}}\overline{c}_1+\left\langle \mathbf v_{1}\right\rangle_{\mathcal{IB}} c_0 \right) +\omega^{\beta-\gamma}\mathcal{K^\star}ac_0^{a-1}\langle c_1\rangle_{\mathcal I\Gamma}=0.
	\end{align}
\end{linenomath*}
Let 
\begin{linenomath*}
	\begin{align}
&\tilde{\textbf{D}}^\star=\omega^{1-\gamma}\left\langle \textbf{D}\nabla_{\mathbf y}\boldsymbol \chi\right\rangle+ \omega^{1-\alpha}\left\langle\boldsymbol\chi \mathbf k\right\rangle\cdot\nabla_{\mathbf x} P_0
\end{align}
\end{linenomath*}
$\tilde{\textbf{D}}^\star$ is a positive definite tensor. \deleted{and $\textbf{D}'$ is a positive  dispersion vector.} Accordingly, \eqref{eq:46} can be written as
\begin{linenomath*}
	\begin{align}\label{eq:49}
\omega\left\langle\dfrac{\partial c_1}{\partial t}\right\rangle_{\mathcal{IB}}&+\omega\left\langle\dfrac{\partial c_2}{\partial \tau}\right\rangle_{\mathcal{IB}} -\phi^{-1} \nabla_{\mathbf{x}}\cdot\left(\tilde{\textbf{D}}^\star\cdot \nabla_{\mathbf x}\left\langle c_0\right\rangle\right) \nonumber \\
&+\omega^{1-\alpha} \nabla_{\mathbf x}\cdot\left(\left\langle  \mathbf v_0 \right\rangle_{\mathcal{IB}}\overline{c}_1+\left\langle \mathbf v_{1}\right\rangle_{\mathcal{IB}} c_0 \right) +\omega^{\beta-\gamma}\mathcal{K^\star}\left(a\omega c_0^{a-1}\langle c_1\rangle_{\mathcal I\Gamma}\right)=0.
	\end{align}
\end{linenomath*}
Calculating $\langle \partial c_\omega/\partial t \rangle_{\mathcal{IB}}$,  while retaining terms up to the second order gives
\begin{linenomath*}
\begin{align}\label{eq:timedifferentiation-expansion}
&\left\langle\dfrac{\partial  c}{\partial t}\right\rangle_{\mathcal{IB}}=\dfrac{\partial c_0}{\partial t}+\left\langle\dfrac{\partial  c_1}{\partial \tau}\right\rangle_{\mathcal{IB}}+\omega\left(\left\langle\dfrac{\partial  c_1}{\partial t}\right\rangle_{\mathcal{IB}}+\left\langle\dfrac{\partial  c_2}{\partial \tau}\right\rangle_{\mathcal{IB}}\right)+\mathcal O(\omega^2).
\end{align}
\end{linenomath*}
where $\left\langle\dfrac{\partial  c}{\partial t}\right\rangle_{\mathcal{IB}}=\dfrac{\partial  \left\langle c\right\rangle_{\mathcal{IB}}}{\partial t}$ because of the Leibniz rule.
Adding \eqref{eq:49} with \eqref{order0integrated} while accounting for \eqref{eq:timedifferentiation-expansion}, yields
\begin{linenomath*}
	\begin{align}\label{eq:a49}
\phi\dfrac{\partial \left\langle c \right\rangle_{\mathcal{IB}}}{\partial t}&= \nabla_{\mathbf{x}}\cdot(\tilde{\textbf{D}}^\star \nabla_{\mathbf x}\left\langle c_0\right\rangle_{\mathcal{IB}}) +\nabla_{\mathbf x}\cdot(\textbf{D}\nabla_{\mathbf x}\left\langle c_0\right\rangle_{\mathcal{IB}})\nonumber \\
&-\omega^{-\alpha} \nabla_{\mathbf x}\cdot\left(\omega\left\langle  \mathbf v_0 \right\rangle\overline{c}_1+\omega\left\langle \mathbf v_{1}\right\rangle c_0 +c_0\langle \mathbf v_0\rangle_{\mathcal{IB}}\right)\nonumber \\
&+\phi\mathcal{K}^{\star}\omega^{\beta-\gamma}(1-c_0^a-a\omega c_0^{a-1}\langle c_1\rangle_{\mathcal I\Gamma}).
	\end{align}
\end{linenomath*}
Since $\overline c_1=\langle c_1\rangle_{\mathcal{IB}}$ and $\langle c_0\rangle_{\mathcal{IB}}\langle \mathbf v_0\rangle=\langle c_0\rangle\langle \mathbf v_0\rangle_{\mathcal{IB}}$, then
\begin{linenomath*}
	\begin{align}
\langle c\rangle_{\mathcal{IB}}\langle \mathbf{v}\rangle=\langle c_0\rangle\langle \mathbf{v}_0\rangle_{\mathcal{IB}}+\omega c_0\langle \mathbf{v}_1\rangle+\omega \overline c_1 \langle \mathbf{v}_0\rangle+\mathcal O(\omega^2).
\end{align}
\end{linenomath*}
%
Assuming  that $\quad\langle\chi\rangle_{\mathcal I\Gamma}\approx\langle\chi\rangle_{\mathcal{IB}}$, then $\langle c_1\rangle_{\mathcal I\Gamma}\approx\langle c_1\rangle_{\mathcal{IB}}$ and
\begin{linenomath*}
\begin{align}
&\langle c_0\rangle_{\mathcal{IB}}^{a}+\omega a\langle c_0\rangle^{a-1}_{\mathcal{IB}}\langle c_1\rangle_{\mathcal{I}\Gamma}\approx\langle c_0\rangle_{\mathcal{IB}}^{a}+\omega a\langle c_0\rangle^{a-1}_{\mathcal{IB}}\langle c_1\rangle_{\mathcal{IB}}=\langle c\rangle^{a}_{\mathcal{IB}}+O(\omega^2).
\end{align}
\end{linenomath*}
Defining
	 \begin{linenomath*}
	 	\begin{align}
	 	&\tilde{\tilde{\textbf{D}}}^\star=\langle\textbf{D}(\mathbf I+\omega^{1-\gamma}\nabla_{\mathbf y}\boldsymbol \chi)\rangle+\omega^{1-\alpha}\langle\boldsymbol\chi \mathbf k\rangle\cdot \nabla_{\mathbf x}P_0,
	 	\end{align}
	 \end{linenomath*}
 \eqref{eq:a49} becomes
\begin{linenomath*}
\begin{align}\label{upscaled}
&\phi\dfrac{\partial \langle c\rangle_{\mathcal{IB}}}{\partial t}=\nabla \cdot (\tilde{\tilde{\textbf{D}}}^\star\nabla\langle c\rangle_{\mathcal{IB}}-\mbox{Pe}\langle c\rangle_{\mathcal{IB}}\langle \mathbf{v}\rangle)+\phi \omega^{-\gamma}\mathcal{K}^{\star}\mbox{Da}(1-\langle c\rangle_{\mathcal{IB}}^a),
\end{align}
\end{linenomath*}
which approximates the space-time average of $c_\omega$ up to an error of order $\omega^2$.

\section{Equations summary}
\subsection{Slowly Fluctuating Regimes: $\varepsilon<\omega$}
\subsubsection{$\mbox{Pe}<1$}
\begin{linenomath*}
	\begin{align}
	&\phi\dfrac{\partial \langle c\rangle_{\mathcal{IB}}}{\partial t}=\nabla\cdot\left[\tilde{\tilde{\textbf{D}}}^\star\nabla\langle c\rangle_{\mathcal{IB}}\right]+\phi \omega^{-\gamma}\mathcal{K}^{\star}\mbox{Da}(1-\langle c\rangle_{\mathcal{IB}}^a),\nonumber
			\end{align}
\end{linenomath*}
with
\begin{linenomath*}
	\begin{align}
	&\tilde{\tilde{\textbf{D}}}^\star=\langle \textbf{D}(\mathbf I+\omega^{1-\gamma}\nabla_{\mathbf y}\boldsymbol{\chi})\rangle
			\end{align}
\end{linenomath*}
\deleted{and $\lambda$} and $\boldsymbol\chi$ defined as the solution of the following boundary value problem in the unit cell $\mathcal B$
\begin{linenomath*}
	\begin{align}
		&\nabla_{\mathbf y}\cdot\textbf{D}(\mathbf I+\omega^{1-\gamma}\nabla_{\mathbf y}\boldsymbol{\chi})=0,  \quad \mbox{subject to} \quad \mathbf n\cdot\textbf{D} (\mathbf I+\omega^{1-\gamma}\nabla_{\mathbf y}\boldsymbol{\chi})=0.
		\end{align}
\end{linenomath*}
%
\subsubsection{$1<\mbox{Pe}< \omega^{-1}$}
\begin{linenomath*}
	\begin{align}
	&\phi\dfrac{\partial \langle c\rangle_{\mathcal{IB}}}{\partial t}=\nabla\cdot\left[\tilde{\tilde{\textbf{D}}}^\star\nabla\langle c\rangle_{\mathcal{IB}}-\mbox{Pe}\langle c\rangle_{\mathcal{IB}}\langle\mathbf v\rangle_{\mathcal{IB}}\right]+\phi \omega^{-\gamma}\mathcal{K}^{\star}\mbox{Da}(1-\langle c\rangle_{\mathcal{IB}}^a), \nonumber
		\end{align}
\end{linenomath*}
with
\begin{linenomath*}
	\begin{align}
&\tilde{\tilde{\textbf{D}}}^\star=\langle \textbf{D}(\mathbf I+\omega^{1-\gamma}\nabla_{\mathbf y}\boldsymbol{\chi})\rangle+\omega^{1-\alpha}\langle\boldsymbol{\chi}\mathbf k\rangle\cdot\nabla_{\mathbf x}P_0,
		\end{align}
\end{linenomath*}
and $\boldsymbol\chi$ defined as the solution of the following boundary value problem in the unit cell $\mathcal B$
\begin{linenomath*}
	\begin{align}
		&\textbf{ D}\nabla^2_{\mathbf y}\lambda=0,  \quad \mbox{subject to} \quad\mathbf n\cdot\textbf{D}\nabla_{\mathbf y}\lambda=0 \mbox{ on }\Gamma,\nonumber\\
		&\nabla_{\mathbf y}\cdot\textbf{D}(\mathbf I+\omega^{1-\gamma}\nabla_{\mathbf y}\boldsymbol{\chi})=0,  \quad \mbox{subject to} \quad \mathbf n\cdot\textbf{D} (\mathbf I+\omega^{1-\gamma}\nabla_{\mathbf y}\boldsymbol{\chi})=0\mbox{ on }\Gamma.
		\end{align}
\end{linenomath*}
\subsection{Moderately Fluctuating Regimes: $\omega^{1/2}<\varepsilon<1$}
\begin{linenomath*}
	\begin{align}
		&\phi\dfrac{\partial \langle c\rangle_{\mathcal{IB}}}{\partial t}=\nabla\cdot\left[\tilde{\tilde{\textbf{D}}}^\star\nabla\langle c\rangle_{\mathcal{IB}}-\mbox{Pe}\langle c\rangle_{\mathcal{IB}}\langle\mathbf v\rangle_{\mathcal{IB}}\right]+\phi \omega^{-\gamma}\mathcal{K}^{\star}\mbox{Da}(1-\langle c\rangle_{\mathcal{IB}}^a), \nonumber
				\end{align}
\end{linenomath*}
with
\begin{linenomath*}
	\begin{align}
		&\tilde{\tilde{\textbf{D}}}^\star=\langle \textbf{D}(\mathbf I+\omega^{1-\gamma}\nabla_{\mathbf y}\boldsymbol{\chi})\rangle+\omega^{1-\alpha}\langle\boldsymbol{\chi}\mathbf k\rangle\cdot\nabla_{\mathbf x}P_0
				\end{align}
\end{linenomath*}
and $\boldsymbol\chi$ defined as the solution of the following boundary value problem in the unit cell $\mathcal B$
\begin{linenomath*}
	\begin{align}				
		&	\omega^{-\alpha}(\mathbf v_0-\langle \mathbf v_0\rangle)-\omega^{-\gamma}\nabla_{\mathbf y}\cdot\textbf{D}(\mathbf I+\omega^{1-\gamma}\nabla_{\mathbf y}\boldsymbol{\chi})+\omega^{1-\gamma-\alpha}\mathbf v_0\cdot(\nabla_{\mathbf y}\boldsymbol{\chi})=0, \nonumber\\ &\quad \mbox{subject to} \quad 
	\mathbf n\cdot\textbf{D} (\mathbf I+\omega^{1-\gamma}\nabla_{\mathbf y}\boldsymbol{\chi})=0, \mbox{ on }\Gamma.
	\end{align}
\end{linenomath*}
\subsection{Higly Fluctuating Regimes: $\varepsilon\gg\omega$}
\subsubsection{$\mbox{Pe}<1$}
\begin{linenomath*}
	\begin{align}
		\phi\dfrac{\partial \langle c\rangle_{\mathcal{IB}}}{\partial t}=\nabla\cdot\left[\tilde{\tilde{\textbf{D}}}^\star\nabla\langle c\rangle_{\mathcal{IB}}\right]+\phi \omega^{-\gamma}\mathcal{K}^{\star}\mbox{Da}(1-\langle c\rangle_{\mathcal{IB}}^a),\nonumber
			\end{align}
\end{linenomath*}
with
\begin{linenomath*}
	\begin{align}
		&\tilde{\tilde{\textbf{D}}}^\star=\langle \textbf{D}(\mathbf I+\omega^{1-\gamma}\nabla_{\mathbf y}\boldsymbol{\chi})\rangle
	\end{align}
\end{linenomath*}		
and $\boldsymbol\chi$ defined as the solution of the following boundary value problem in the unit cell $\mathcal B$
\begin{linenomath*}
	\begin{align}		
			&\dfrac{\partial\boldsymbol{\chi}}{\partial \tau}-\omega^{-\gamma}\nabla_{\mathbf y}\cdot\textbf{D}(\mathbf I+\omega^{1-\gamma}\nabla_{\mathbf y}\boldsymbol{\chi})+	\omega^{-\alpha}(\mathbf v_0-\langle \mathbf v_0\rangle)=0
				\end{align}
\end{linenomath*}
subject to
\begin{linenomath*}
	\begin{align}	
& -\mathbf n\cdot\textbf{D} (\mathbf I+\omega^{1-\gamma}\nabla_{\mathbf y}\boldsymbol{\chi})=0\quad \mbox{on} \quad \mathcal{\varGamma},\nonumber\\
& \boldsymbol\chi(\mathbf y, \tau=0)=\boldsymbol\chi_{\tiny\mbox{in}}(\mathbf y)=0.
	\end{align}
\end{linenomath*}
\subsubsection{$1<\mbox{Pe}< \omega^{-1}$}
\begin{linenomath*}
\begin{align}
&\phi\dfrac{\partial \langle c\rangle_{\mathcal{IB}}}{\partial t}=\nabla\cdot\left[\tilde{\tilde{\textbf{D}}}^\star\nabla\langle c\rangle_{\mathcal{IB}}-\mbox{Pe}\langle c\rangle_{\mathcal{IB}}\langle\mathbf v\rangle_{\mathcal{IB}}\right]+\phi \omega^{-\gamma}\mathcal{K}^{\star}\mbox{Da}(1-\langle c\rangle_{\mathcal{IB}}^a), \nonumber
\end{align}
\end{linenomath*}
with
\begin{linenomath*}
	\begin{align}
			&\tilde{\tilde{\textbf{D}}}^\star=\langle \textbf{D}(\mathbf I+\omega^{1-\gamma}\nabla_{\mathbf y}\boldsymbol{\chi})\rangle+\omega^{1-\alpha}\langle\boldsymbol{\chi}\mathbf k\rangle\cdot\nabla_{\mathbf x}P_0
							\end{align}
\end{linenomath*}
 and $\boldsymbol\chi$ defined as the solution of the following boundary value problem in the unit cell $\mathcal B$
\begin{linenomath*}
	\begin{align}
				&\dfrac{\partial \chi}{\partial \tau}+\omega^{-\alpha}(\mathbf v_0-\langle\mathbf v_0\rangle)+\omega^{1-\gamma-\alpha}\mathbf v_0\cdot(\nabla_{\mathbf y}\boldsymbol{\chi})=0, \nonumber
	\end{align}
\end{linenomath*}
subject to
\begin{linenomath*}
	\begin{align}	
& -\mathbf n\cdot\textbf{D} (\mathbf I+\omega^{1-\gamma}\nabla_{\mathbf y}\boldsymbol{\chi})=0\quad \mbox{on} \quad \mathcal{\varGamma},\nonumber\\
& \boldsymbol\chi(\mathbf y, \tau=0)=\boldsymbol\chi_{\tiny\mbox{in}}(\mathbf y)=0.
	\end{align}
\end{linenomath*}

\section{Nomenclature}
\begin{linenomath*}
\begin{align}
& \mathcal{B}:  \mbox{Pore-scale domain in the unit cell} \quad Y \nonumber\\
& \mathcal{I}: \mbox{Temporal unit cell} \nonumber\\
&c_{\omega}:  \mbox{Dimensionless pore-scale concentration} \nonumber\\
& c_{\mbox{\tiny{in}}}(\mathbf x):  \mbox{Dimensionless initial pore-scale concentration} \nonumber\\
&c_{D}(t): \mbox{Dimensionless time-varying concentration at a  Dirichlet boundary $\partial  \Omega_D$}  \nonumber\\
& \langle c\rangle_{\mathcal{IB}}:  \mbox{Average of pore-scale concentration over the pore volume $\mathcal{B}$ and the time interval $\mathcal{I}$} \nonumber\\
& \langle c\rangle:  \mbox{Average of pore-scale concentration over the unit cell $Y$ and the time interval $\mathcal{I}$, such that  $\langle c\rangle=\phi \langle c\rangle_{\mathcal{IB}}$} \nonumber\\
& \mathbf{D}: \mbox{Dimensionless molecular diffusion coefficient} \nonumber\\
&\mbox{Da}: \mbox{Damk\"{o}hler number} \nonumber \\
&\mbox{Pe}: \mbox{Pecl\'{e}t number} \nonumber\\
&l: \mbox{Characteristic length of periodic unit cell $Y$} \nonumber\\
&L: \mbox{Characteristic length of the macroscopic porous medium domain $\Omega$} \nonumber\\
& a: \mbox{Order of the heterogeneous reaction} \nonumber\\
&\hat{p}: \mbox{Dimensional dynamic pressure} \nonumber\\
& \mu: \mbox{Dynamic viscosity of the fluid} \nonumber\\
& \varepsilon=\dfrac{l}{L}: \mbox{Spatial scale separation parameter} \nonumber\\
&\omega=\dfrac{\hat \tau}{T}: \mbox{Temporal scale separation parameter} \nonumber\\
&\phi: \mbox{Unit cell porosity}\nonumber\\
&\hat{\Omega}: \mbox{Porous medium domain}\nonumber\\
&\hat{\Omega}_p: \mbox{Volume of the pore phase in $\hat\Omega$}\nonumber\\
&\hat{\Omega}_s: \mbox{Volume of the solid phase  in $\hat\Omega$}\nonumber\\
&\partial\hat{\Omega}: \mbox{Outer boundary of the porous medium $\hat\Omega$}\nonumber\\
&\hat\Gamma: \mbox{Boundary between solid and pore phase}\nonumber\\
&\hat{\mathbf{v}}_{\varepsilon}: \mbox{Dimensional pore-scale velocity}\nonumber\\
&\boldsymbol\chi: \mbox{Closure variable in the unit cell}\nonumber\\
&\hat{Y}: \mbox{Unit cell domain}\nonumber\\
&\hat{\mathcal{B}}: \mbox{Solid phase in the unit cell domain $Y$}\nonumber\\
&\hat{\mathcal G}: \mbox{Pore phase in the unit cell domain $Y$}\nonumber
\end{align}
\end{linenomath*}
\begin{linenomath*}
\begin{align}
&\mathbf x: \mbox{Slow spatial scale}\nonumber\\
&t: \mbox{Slow time scale}\nonumber\\
&\mathbf y: \mbox{Fast spatial scale}\nonumber\\
&\tau: \mbox{Fast time scale}\nonumber\\
&U: \mbox{Characteristic velocity}\nonumber\\
&p: \mbox{Dimensionless pressure}\nonumber\\
&\hat{t}_{\mbox{\tiny{d,micro}}}: \mbox{Dimensional time-scale for diffusion at microscale}\nonumber\\
&\hat{t}_{\mbox{\tiny{d,macro}}}: \mbox{Dimensional time-scale for diffusion at microscale}\nonumber\\
&\hat{t}_{\mbox{\tiny{a,micro}}}: \mbox{Dimensional time-scale for advection at microscale}\nonumber\\
&\hat{t}_{\mbox{\tiny{a,macro}}}: \mbox{Dimensional time-scale for advection at macroscale}\nonumber\\
&\tau_c=\dfrac{L^2}{D}: \mbox{Characteristic time}\nonumber\\
&T: \mbox{Observation time-scale}\nonumber\\
&\hat{\tau}_a: \mbox{Advection time-scale}\nonumber\\
&\hat{\tau}_d: \mbox{Diffusion time-scale}\nonumber\\
&\hat{\tau}_r: \mbox{Reaction time-scale}\nonumber\\
&\hat{k}: \mbox{Dimensional pore-scale heterogeneous reaction rate}\nonumber\\
&\gamma: \mbox{The parameter connecting spatial and temporal scale separation parameters}\nonumber\\
&\psi_\varepsilon: \mbox{Any arbitrary pore-scale quantity}\nonumber\\
&\alpha: \mbox{Parameter defining Pecl\'{e}t, $\mbox{Pe}=\omega^{-\alpha}$}\nonumber\\
&\beta: \mbox{Parameter defining Damk\"{o}hler, $\mbox{Da}=\omega^{\beta}$}\nonumber\\
&\mathbf K: \mbox{Dimensionless permeability tensor}\nonumber\\
&\mathbf k: \mbox{Closure variable}\nonumber\\
&\mathbf{a}: \mbox{Closure variable}\nonumber\\
&\mathcal{K^\star}: \mbox{Effective reaction rate}\nonumber\\
&\tilde{\tilde{\textbf{D}}}^\star: \mbox{Effective dispersion tensor}\nonumber\\
&\nabla_x P_0: \mbox{Macroscopic pressure gradient}\nonumber\\
&\mathbf n: \mbox{Unit vector normal to the boundary}\nonumber\\
&c_0,c_1,c_2,...: \mbox{Expansions of pore-scale concentration}\nonumber\\
&\mathbf{v}_0,\mathbf{v}_1,\mathbf{v}_2,\cdots: \mbox{Expansions of pore-scale velocity}\nonumber
\end{align}
\end{linenomath*}

\begin{acknowledgments}
Financial support for this work was provided by the Stanford University Petroleum Research Institute (SUPRI-B Industrial Affiliates Program). The Author is grateful to Professor Hamdi Tchelepi from the Energy Resources Engineering Department at Stanford University for reviewing the content of this paper and providing valuable feedback. The author declares no known competing financial interests or personal relationships that could have appeared to influence the work reported in this paper.
\end{acknowledgments}
%

\bibliography{Manuscript}

\begin{thebibliography}{74}
\providecommand{\natexlab}[1]{#1}
\expandafter\ifx\csname urlstyle\endcsname\relax
  \providecommand{\doi}[1]{doi:\discretionary{}{}{}#1}\else
  \providecommand{\doi}{doi:\discretionary{}{}{}\begingroup
  \urlstyle{rm}\Url}\fi

\bibitem[{\textit{Abraham et~al.}(1998)\textit{Abraham, Broughton, Bernstein,
  and Kaxiras}}]{Abraham-1998-Spanning}
Abraham, F.~F., J.~Q. Broughton, N.~Bernstein, and E.~Kaxiras (1998), Spanning
  the length scales in dynamic simulations, \textit{Comput. Phys.},
  \textit{12}(538).

\bibitem[{\textit{Acharya et~al.}(2005)\textit{Acharya, Van~der Zee, and
  Leijnse}}]{acharya2005transport}
Acharya, R.~C., S.~E. A. T.~M. Van~der Zee, and A.~Leijnse (2005), Transport
  modeling of nonlinearly adsorbing solutes in physically heterogeneous pore
  networks, \textit{Water Resour. Res.}, \textit{41}(2).

\bibitem[{\textit{Alexander et~al.}(2002)\textit{Alexander, Garcia, and
  Tartakovsky}}]{alexander-2002-algorithm}
Alexander, F.~J., A.~L. Garcia, and D.~M. Tartakovsky (2002), Algorithm
  refinement for stochastic partial differential equations: 1. {L}inear
  diffusion, \textit{J. Comput. Phys.}, \textit{182}, 47--66.

\bibitem[{\textit{Alexander et~al.}(2005)\textit{Alexander, Garcia, and
  Tartakovsky}}]{Alexander-2005-noise}
Alexander, F.~J., A.~L. Garcia, and D.~M. Tartakovsky (2005), Noise in
  algorithm refinement methods, \textit{Comput. Sci. Eng.}, \textit{7}(3),
  32--38.

\bibitem[{\textit{Allaire et~al.}(2010)\textit{Allaire, Mikelic, and
  Piatnitski}}]{allaire2010homogenization}
Allaire, G., A.~Mikelic, and A.~Piatnitski (2010), Homogenization approach to
  the dispersion theory for reactive transport through porous media,
  \textit{SIAM J. Math. Anal.}, \textit{42}(1), 125--144.

\bibitem[{\textit{Arbogast et~al.}(2007)\textit{Arbogast, Pencheva, Wheeler,
  and Yotov}}]{arbogast-2007-multiscale}
Arbogast, T., G.~Pencheva, M.~F. Wheeler, and I.~Yotov (2007), A multiscale
  mortar mixed finite element method, \textit{Multiscale Model. Simul.},
  \textit{6}(1), 319--346.

\bibitem[{\textit{Auriault}(1991)}]{Auriault-1991-heteronenous}
Auriault, J.~L. (1991), Heterogenous medium. is an equivalent macroscopic
  description possible?, \textit{Int. J. Engng Sci.}, \textit{29}(7), 785--795.

\bibitem[{\textit{Auriault}(2019)}]{auriault2019comments}
Auriault, J.-L. (2019), Comments on the paper “theory and applications of
  macroscale models in porous media” by ilenia battiato et al,
  \textit{Transport in Porous Media}, \textit{130}(2), 611--612.

\bibitem[{\textit{Auriault and Adler}(1995)}]{auriault1995taylor}
Auriault, J.-L., and P.~M. Adler (1995), Taylor dispersion in porous media:
  analysis by multiple scale expansions, \textit{Adv. Water Resour.},
  \textit{18}(4), 217--226.

\bibitem[{\textit{Beese and Wierenga}(1980)}]{Beese-1980-solute}
Beese, F., and P.~J. Wierenga (1980), Solute transport through soil with
  adsorption and root water uptake computed witha transient and a
  constant-flux, \textit{Soil Sci.}, \textit{129}(245).

\bibitem[{\textit{Bensoussan et~al.}(1978)\textit{Bensoussan, Lions, and
  Papanicolaou}}]{bensoussan1978asymptotic}
Bensoussan, A., J.-L. Lions, and G.~Papanicolaou (1978), \textit{Asymptotic
  analysis for periodic structures}, vol.~5, North-Holland Publishing Company
  Amsterdam.

\bibitem[{\textit{Bogers et~al.}(2013)\textit{Bogers, Kumar, Notten,
  Oudenhoven, and Pop}}]{Bogers-2013-multiscale}
Bogers, J., K.~Kumar, P.~H.~L. Notten, J.~F.~M. Oudenhoven, and I.~S. Pop
  (2013), A multiscale domain decomposition approach for chemical vapor
  deposition, \textit{J. Comput. Appl. Math.}, \textit{246}, 65--73.

\bibitem[{\textit{Brenner}(1980)}]{brenner1980dispersion}
Brenner, H. (1980), Dispersion resulting from flow through spatially periodic
  porous media, \textit{Philos. T. Roy. Soc. A}, \textit{297}(1430), 81--133.

\bibitem[{\textit{Brenner}(1987)}]{brenner1986transport}
Brenner, H. (1987), \textit{Transport Processes in Porous Media}, McGraw-Hill.

\bibitem[{\textit{Bresler and Dagan}(1982)}]{Bresler-1982-Unsaturated}
Bresler, E., and G.~Dagan (1982), Unsaturated flow in spatially variable
  fields: 3. {S}olute transport models and their application to two fields,
  \textit{Water Resour. Res.}, \textit{19}, 429--435.

\bibitem[{\textit{Bringedal et~al.}(2016)\textit{Bringedal, Berre, Pop, and
  Radu}}]{Bringedal-2016-upscaling}
Bringedal, C., I.~Berre, I.~S. Pop, and F.~A. Radu (2016), Upscaling of
  nonisothermal reactive porous media flow under dominant {P}\'eclet number:
  {T}he effect of cganging porosity, \textit{SIAM Multiscale Model Simul.},
  \textit{14}(1), 502--533.

\bibitem[{\textit{Cushman et~al.}(2002)\textit{Cushman, Bennethum, and
  Hu}}]{cushman2002primer}
Cushman, J.~H., L.~S. Bennethum, and B.~X. Hu (2002), A primer on upscaling
  tools for porous media, \textit{Adv. Water Resour.}, \textit{25}(8),
  1043--1067.

\bibitem[{\textit{Danckwerts}(1953)}]{Danckwerts-1953-Continuous}
Danckwerts, P.~V. (1953), Continuous flow systems: Distribution of residence
  times, \textit{Chem. Eng. Sci.}, \textit{2}, 1--13.

\bibitem[{\textit{Davit and Quintard}(2012)}]{davit-2012-comment}
Davit, Y., and M.~Quintard (2012), {C}omment on `{F}requency-dependent
  dispersion in porous media', \textit{Phys. Rev. E}, \textit{86}(013201).

\bibitem[{\textit{Davit et~al.}(2013)\textit{Davit, Bell, Byrne, Chapman,
  Kimpton, Lang, Leonard, Oliver, Pearson, Shipley
  et~al.}}]{davit2013homogenization}
Davit, Y., C.~G. Bell, H.~M. Byrne, L.~A.~C. Chapman, L.~S. Kimpton, G.~E.
  Lang, K.~H.~L. Leonard, J.~M. Oliver, N.~C. Pearson, R.~J. Shipley, et~al.
  (2013), Homogenization via formal multiscale asymptotics and volume
  averaging: How do the two techniques compare?, \textit{Adv. Water Resour.},
  \textit{62}, 178--206.

\bibitem[{\textit{Dentz and Carrera}(2003)}]{dentz2003effective}
Dentz, M., and J.~Carrera (2003), Effective dispersion in temporally
  fluctuating flow through a heterogeneous medium, \textit{Phys. Rev. E},
  \textit{68}(3), 036,310.

\bibitem[{\textit{Fish and Chen}(2004)}]{fish2004space}
Fish, J., and W.~Chen (2004), Space--time multiscale model for wave propagation
  in heterogeneous media, \textit{Comput. Method Appl. M.}, \textit{193}(45),
  4837--4856.

\bibitem[{\textit{Flekkoy et~al.}(2000)\textit{Flekkoy, Wagner, and
  Feder}}]{Flekkoy-2000-hybrid}
Flekkoy, E.~G., G.~Wagner, and J.~Feder (2000), Hybrid model for combined
  particle and continuum dynamics, \textit{Europhys. Lett.}, \textit{52}(271).

\bibitem[{\textit{Ganis et~al.}(2014)\textit{Ganis, Juntunen, Pencheva,
  Wheeler, and Yotov}}]{ganis2014global}
Ganis, B., M.~Juntunen, G.~Pencheva, M.~F. Wheeler, and I.~Yotov (2014), A
  global jacobian method for mortar discretizations of nonlinear porous media
  flows, \textit{SIAM J. Sci. Comput}, \textit{36}(2), A522--A542.

\bibitem[{\textit{Gray and Miller}(2005)}]{Gray-2005-Thermodynamically}
Gray, W.~G., and C.~T. Miller (2005), Thermodynamically constrained averaging
  theory approach for modeling flow and transport phenomena in porous medium
  systems: 1. motivation and overview, \textit{Adv. Water Resour.},
  \textit{28}(2), 160--180.

\bibitem[{\textit{Gray and Miller}(2014)}]{gray-2014-intro}
Gray, W.~G., and C.~T. Miller (2014), \textit{Introduction to the
  Thermodynamically Constrained Averaging Theory for Porous Medium Systems -
  Advances in Geophysical and Environmental Mechanics and Mathematics},
  Springer International Publishing.

\bibitem[{\textit{Hadjiconstantinou and
  Patera}(1997)}]{Hadjiconstantinou-1997-heterogeneous}
Hadjiconstantinou, N., and A.~Patera (1997), Heterogenous atomistic-continuum
  representations for dense fluid systems, \textit{Int. J. Mod. Phys. C},
  \textit{8}(967).

\bibitem[{\textit{He and Sykes}(1996)}]{he-1996-spatial}
He, Y., and J.~F. Sykes (1996), On the spatial-temporal averaging method for
  modeling transport in porous media, \textit{Transp. Porous Media},
  \textit{22}, 1--51.

\bibitem[{\textit{Helming et~al.}(2013)\textit{Helming, Flemisch, Wolff,
  Ebigbo, and Class}}]{helming-2013-model}
Helming, R., B.~Flemisch, M.~Wolff, A.~Ebigbo, and H.~Class (2013), Model
  coupling for multiphase flow in porous media, \textit{Adv. Water Resour.},
  \textit{51}, 52--66.

\bibitem[{\textit{Hornung}(2012)}]{hornung2012homogenization}
Hornung, U. (2012), \textit{Homogenization and porous media}, vol.~6, Springer
  Science \& Business Media.

\bibitem[{\textit{Hornung et~al.}(1994)\textit{Hornung, J\"{a}ger, and
  Mikeli\'{c}}}]{hornung1994reactive}
Hornung, U., W.~J\"{a}ger, and A.~Mikeli\'{c} (1994), Reactive transport
  through an array of cells with semi-permeable membranes,
  \textit{RAIRO-Mod\'{e}lisation math\'{e}matique et analyse num\'{e}rique},
  \textit{28}(1), 59--94.

\bibitem[{\textit{Kumar et~al.}(2011)\textit{Kumar, van Noorden, and
  Pop}}]{Kumar-2011-effective}
Kumar, K., T.~L. van Noorden, and I.~S. Pop (2011), Effective disperion
  equations for reactive flows involving free boundaries at the microscale,
  \textit{SIAM Multiscale Model Simul.}, \textit{9}(1), 29--58.

\bibitem[{\textit{Kumar et~al.}(2014)\textit{Kumar, {van N}oorden, and
  Pop}}]{Kumar-2014-Upscaling}
Kumar, K., T.~{van N}oorden, and I.~S. Pop (2014), Upscaling of reactive flows
  in domains with moving oscilating boundaries, \textit{Discrete Contin. Dyn.
  Syst. Ser. S}, \textit{7}(1), 95--111.

\bibitem[{\textit{Mehmani and Balhoff}(2014)}]{mehmani2014bridging}
Mehmani, Y., and M.~T. Balhoff (2014), Bridging from pore to continuum: A
  hybrid mortar domain decomposition framework for subsurface flow and
  transport, \textit{SIAM Multiscale Model. Sim.}, \textit{12}(2), 667--693.

\bibitem[{\textit{Mehmani et~al.}(2012)\textit{Mehmani, Sun, Balhoff, Eichhubl,
  and Bryant}}]{mehmani2012multiblock}
Mehmani, Y., T.~Sun, M.~T. Balhoff, P.~Eichhubl, and S.~Bryant (2012),
  Multiblock pore-scale modeling and upscaling of reactive transport:
  Application to carbon sequestration, \textit{Transp. Porous Med.},
  \textit{95}(2), 305--326.

\bibitem[{\textit{Mikelic et~al.}(2006)\textit{Mikelic, Devigne, and
  Van~Duijn}}]{mikelic2006rigorous}
Mikelic, A., V.~Devigne, and C.~J. Van~Duijn (2006), Rigorous upscaling of the
  reactive flow through a pore, under dominant peclet and damkohler numbers,
  \textit{SIAM J. Math. Anal.}, \textit{38}(4), 1262--1287.

\bibitem[{\textit{Miller et~al.}(2013)\textit{Miller, Dawson, Farthing, Hou,
  Huang, Kees, Kelley, and Langtangen}}]{miller-2013-Numerical}
Miller, C.~T., C.~N. Dawson, M.~W. Farthing, T.~Y. Hou, J.~Huang, C.~E. Kees,
  C.~T. Kelley, and H.~P. Langtangen (2013), Numerical simulation of water
  resources problems: models, methods and trends, \textit{Adv. Water Resour.},
  \textit{51}, 405--437.

\bibitem[{\textit{Moyne}(1997)}]{moyne-1997-two}
Moyne, C. (1997), Two-equation model for a diffusive process in porous media
  using the volume averaging method with an unsteady-state closure,
  \textit{Adv. Water Resour.}, \textit{20}(2-3), 63--76.

\bibitem[{\textit{Nissan et~al.}(2017)\textit{Nissan, Dror, and
  Berkowitz}}]{Nissan-2017-time}
Nissan, A., I.~Dror, and B.~Berkowitz (2017), Time dependent velocity field
  controls on anomalous chemical transport in porous media, \textit{Water
  Resour. Res.}, \textit{53}(5), 3760--3769.

\bibitem[{\textit{Pavliotis}(2002)}]{pavliotis2002homogenization}
Pavliotis, G.~A. (2002), Homogenization theory for advection diffusion
  equations with mean flow, Ph.D. thesis, Rensselaer Polytechnic Institute.

\bibitem[{\textit{Pavliotis and Kramer}(2002)}]{pavliotis2002homogenized}
Pavliotis, G.~A., and P.~R. Kramer (2002), Homogenized transport by a
  spatiotemporal mean flow with small-scale periodic fluctuations, in
  \textit{Proc. of the IV International Conference on Dynamical Systems and
  Differential Equations, May}, pp. 24--27.

\bibitem[{\textit{Pavliotis and Stuart}(2008)}]{pavliotis2008multiscale}
Pavliotis, G.~A., and A.~Stuart (2008), \textit{Multiscale methods: averaging
  and homogenization}, Springer Science \& Business Media.

\bibitem[{\textit{Peszy\'{n}ska et~al.}(2002)\textit{Peszy\'{n}ska, Wheeler,
  and Yotov}}]{Malgo-2002-mortar}
Peszy\'{n}ska, M., M.~F. Wheeler, and I.~Yotov (2002), Mortar upscaling for
  multiphase flow in porous media, \textit{Comput. Geosci.}, \textit{6},
  73--100.

\bibitem[{\textit{Pool et~al.}(2014)\textit{Pool, Post, and
  Simmons}}]{pool2014effects}
Pool, M., V.~E. Post, and C.~T. Simmons (2014), Effects of tidal fluctuations
  on mixing and spreading in coastal aquifers: Homogeneous case, \textit{Water
  Resour. Res.}, \textit{50}(8), 6910--6926.

\bibitem[{\textit{Pool et~al.}(2015)\textit{Pool, Post, and
  Simmons}}]{pool2015effects}
Pool, M., V.~E.~A. Post, and C.~T. Simmons (2015), Effects of tidal
  fluctuations and spatial heterogeneity on mixing and spreading in spatially
  heterogeneous coastal aquifers, \textit{Water Resour. Res.}, \textit{51}(3),
  1570--1585.

\bibitem[{\textit{Pool et~al.}(2016)\textit{Pool, Dentz, and
  Post}}]{pool2016transient}
Pool, M., M.~Dentz, and V.~E. Post (2016), Transient forcing effects on mixing
  of two fluids for a stable stratification, \textit{Water Resour. Res.},
  \textit{52}(9), 7178--7197.

\bibitem[{\textit{Pope}(2000)}]{pope}
Pope, S.~B. (2000), \textit{Turbulent flows}, Cambridge University Press,
  Cambridge, NY.

\bibitem[{\textit{Rajabi}(2021)}]{rajabi2021stochastic}
Rajabi, F. (2021), Stochastic models for nonlinear transport in multiphase and
  multiscale heterogeneous media, Ph.D. thesis, Stanford University.

\bibitem[{\textit{Rajabi and Battiato}(2015)}]{rajabi2015spatio}
Rajabi, F., and I.~Battiato (2015), Spatio-temporal upscaling of reactive
  transport in porous media for ultra-long time predictions: Theory and
  numerical experiments, in \textit{AGU Fall Meeting Abstracts}, vol. 2015, pp.
  H51F--1434.

\bibitem[{\textit{Rajabi and Battiato}(2017)}]{rajabi2017frequency}
Rajabi, F., and I.~Battiato (2017), Frequency dependent macro-dispersion
  induced by oscillatory inputs and spatial heterogeneity, in \textit{AGU Fall
  Meeting Abstracts}, vol. 2017, pp. H11G--1276.

\bibitem[{\textit{Roubinet and Tartakovsky}(2013)}]{roubinet2013hybrid}
Roubinet, D., and D.~M. Tartakovsky (2013), Hybrid modeling of heterogeneous
  geochemical reactions in fractured porous media, \textit{Water Resour. Res.},
  \textit{49}(12), 7945--7956.

\bibitem[{\textit{Russo et~al.}(1989)\textit{Russo, Jury, and
  Butters}}]{Russo-1989-Numerical}
Russo, D., W.~A. Jury, and G.~L. Butters (1989), Numerical analysis of solute
  transport during transient irrigation: 1. {T}he effect of hysterisis and
  profile heterogeneity, \textit{Water Resour. Res.}, \textit{25}, 2109--2128.

\bibitem[{\textit{Shapiro and Brenner}(1988)}]{shapiro1988dispersion}
Shapiro, M., and H.~Brenner (1988), Dispersion of a chemically reactive solute
  in a spatially periodic model of a porous medium, \textit{Chemical
  engineering science}, \textit{43}(3), 551--571.

\bibitem[{\textit{Shenoy et~al.}(1999)\textit{Shenoy, Miller, Tadmor, Rodney,
  Phillips, and Ortiz}}]{Shenoy-1999-adaptive}
Shenoy, V.~B., R.~Miller, E.~B. Tadmor, D.~Rodney, R.~Phillips, and M.~Ortiz
  (1999), An adaptive finite element approach to atomic-scale mechanics-the
  quasicontinuum method, \textit{J. Mech. Phys. Solids}, \textit{47}(611).

\bibitem[{\textit{Smith}(1981)}]{Smith-1981-delay}
Smith, R. (1981), A delay-diffusion description for contaminnat dispersion,
  \textit{J. Fluid Mech.}, \textit{105}, 469--486.

\bibitem[{\textit{Smith}(1982)}]{smith-1982-contaminant}
Smith, R. (1982), Contaminant dispersion in oscillatory flows, \textit{J. Fluid
  Mech.}, \textit{114}, 379--398.

\bibitem[{\textit{Stegen et~al.}(2016)\textit{Stegen, Fredickson, Wilkins,
  Konopa, /c. Nelson, Arntzen, Chrisler, Chu, Danczak, Fansler, Kennedy, Resch,
  and Tfaily}}]{Stegen-2016-Groundwater}
Stegen, J.~C., J.~K. Fredickson, M.~J. Wilkins, A.~E. Konopa, W.~/c. Nelson,
  E.~V. Arntzen, W.~B. Chrisler, R.~Chu, R.~E. Danczak, S.~J. Fansler, D.~W.
  Kennedy, C.~T. Resch, and M.~M. Tfaily (2016), Groundwater-surface water
  mixing shifts ecological assembly processes and stimulates organic carbon
  turnover, \textit{Nature Commun.}, \textit{7}(11237).

\bibitem[{\textit{Tartakovsky et~al.}(2008)\textit{Tartakovsky, Tartakovsky,
  Scheibe, and Meakin}}]{tartakovsky2008hybrid}
Tartakovsky, A.~M., D.~M. Tartakovsky, T.~D. Scheibe, and P.~Meakin (2008),
  Hybrid simulations of reaction-diffusion systems in porous media,
  \textit{SIAM J. Sci. Comput.}, \textit{30}(6), 2799--2816.

\bibitem[{\textit{Tartakovsky}(2013)}]{tartakovsky-2013-assessment}
Tartakovsky, D.~M. (2013), Assessment and management of risk in subsurface
  hydrology: {A} review and pers[ective, \textit{Adv. Water Resour.},
  \textit{51}, 247--260.

\bibitem[{\textit{Taverniers and Tartakovsky}(2017)}]{Taverniers-2017-tightly}
Taverniers, S., and D.~M. Tartakovsky (2017), A tightly-coupled
  domain-decomposition approach for highly nonlinear stochastic multiphysics
  systems, \textit{J. Comput. Phys.}, \textit{330}, 884--901.

\bibitem[{\textit{Taylor}(1953)}]{taylor1953dispersion}
Taylor, G. (1953), Dispersion of soluble matter in solvent flowing slowly
  through a tube, in \textit{P. Roy. Soc. Lond. A Mat.}, vol. 219, pp.
  186--203, The Royal Society.

\bibitem[{\textit{Taylor}(1959)}]{Taylor-1959-present}
Taylor, G.~I. (1959), The present position in the theory of turbulent
  diffusion, \textit{Adv. Geophys.}, \textit{6}, 101--112.

\bibitem[{\textit{Tiwari and Klar}(1998)}]{Tiwari-1998-coupling}
Tiwari, S., and A.~Klar (1998), Coupling of the {B}oltzmann and {E}uler
  equations with adaptive domain decomposition procedure, \textit{J. Comput.
  Phys.}, \textit{144}(710).

\bibitem[{\textit{Valdes-{P}arada and {A}lvarez
  {R}amirez}(2011)}]{parada-2011-Frequency}
Valdes-{P}arada, F.~J., and J.~{A}lvarez {R}amirez (2011), Frequency-dependent
  dispersion in porous media, \textit{Phys. Rev. E}, \textit{84}(031201).

\bibitem[{\textit{Valdes-{P}arada and {A}lvarez
  {R}amirez}(2012)}]{parada-2011-reply}
Valdes-{P}arada, F.~J., and J.~{A}lvarez {R}amirez (2012), Reply to ``{C}omment
  on `{F}requency-dependent dispersion in porous media''', \textit{Phys. Rev.
  E}, \textit{86}(013202).

\bibitem[{\textit{van Noorden et~al.}(2010)\textit{van Noorden, Popo, Ebigbo,
  and Helming}}]{Noorden-2010-upscaled}
van Noorden, T.~L., I.~S. Popo, A.~Ebigbo, and R.~Helming (2010), An upscaled
  model for biofilm growth in a thin strip, \textit{Water Resour. Res.},
  \textit{46}(W06505).

\bibitem[{\textit{Wadsworth and Erwin}(1990)}]{Wadsworth-1990-One}
Wadsworth, D.~C., and D.~A. Erwin (1990), One-dimensional hybrid
  continuum/particle simulation approach for rarefied hypersonic flows,
  \textit{AIAA Paper}, \textit{90-1690}.

\bibitem[{\textit{Wang et~al.}(2009)\textit{Wang, Quinland, and
  Tartakovsky}}]{wang-2009-effects}
Wang, P., P.~Quinland, and D.~M. Tartakovsky (2009), Effects of spatio-temporal
  variability of precipitation on contaminant migration in the vadose zone,
  \textit{Geophys. Res. Lett.}, \textit{36}(L12404).

\bibitem[{\textit{Whitaker}(1999)}]{whitaker2013method}
Whitaker, S. (1999), \textit{The method of volume averaging}, vol.~13, Springer
  Science \& Business Media.

\bibitem[{\textit{Wood}(2009)}]{wood2009role}
Wood, B.~D. (2009), The role of scaling laws in upscaling, \textit{Adv. Water
  Resour.}, \textit{32}(5), 723--736.

\bibitem[{\textit{Wood and Valdes-Parada}(2013)}]{wood2013volume}
Wood, B.~D., and F.~J. Valdes-Parada (2013), Volume averaging: local and
  nonlocal closures using a green's function approach, \textit{Adv. Water
  Resour.}, \textit{51}, 139--167.

\bibitem[{\textit{Wood et~al.}(2003)\textit{Wood, Cherblanc, Quintard, and
  Whitaker}}]{wood2003volume}
Wood, B.~D., F.~Cherblanc, M.~Quintard, and S.~Whitaker (2003), Volume
  averaging for determining the effective dispersion tensor: Closure using
  periodic unit cells and comparison with ensemble averaging, \textit{Water
  Resour. Res.}, \textit{39}(8).

\bibitem[{\textit{Yin et~al.}(2015)\textit{Yin, f.~sykes, and
  Normani}}]{yin-2015-impacts}
Yin, Y., J.~f.~sykes, and S.~D. Normani (2015), Imoacts of spatial and temporal
  recharge on field-scale contaminant transport model calibration, \textit{J.
  Hydrol.}, \textit{527}, 77--87.

\bibitem[{\textit{Yousefzadeh}(2020)}]{yousefzadeh2020numerical}
Yousefzadeh, M. (2020), \textit{Numerical Simulation of Fluid-Mineral
  Interaction and Reactive Transport in Porous and Fractured Media}, Stanford
  University.

\end{thebibliography}

\newpage

\listofchanges

\end{document}